\documentclass[conference,compsoc]{IEEEtran}

\usepackage[skip=\baselineskip]{caption}
\usepackage{xspace}
\usepackage{pifont}
\usepackage{enumitem}
\usepackage{multirow}
\usepackage{xcolor}
\usepackage{tcolorbox}
\usepackage{subfigure}
\usepackage{booktabs}
\usepackage{amssymb}
\usepackage[hyphens]{url}
\usepackage[nospace]{cite}
\definecolor{customblue}{HTML}{006ca6}
\definecolor{customgreen}{HTML}{009264}
\definecolor{custombrown}{HTML}{b80d57}
\AtEndPreamble{
\usepackage{hyperref}
 \hypersetup{
  colorlinks = true,
  linkcolor = customblue,
  anchorcolor = purple,
  citecolor = customgreen,
  filecolor = purple,
  urlcolor = black
 }
}

\begin{document}

\title{A Big Step Forward? A User-Centric Examination of \\iOS App Privacy Report and Enhancements}

\author{
\IEEEauthorblockN{Liu Wang$^{1}$, Dong Wang$^{1}$, Shidong Pan$^{2}$, Zheng Jiang$^{1}$, Haoyu Wang$^{3}$, and Yi Wang$^{1}$}
\IEEEauthorblockA{$^{1}$Beijing University of Posts and Telecommunications, China  $^{2}$Columbia University\\
$^{3}$Huazhong University of Science and Technology, China\\
$^{1}$\{w\_liu,wangdong,jiangzheng,yiwang\}@bupt.edu.cn, $^{2}$sp4471@columbia.edu, $^{3}$haoyuwang@hust.edu.cn}
\thanks{\IEEEauthorrefmark{1}Yi Wang (yiwang@bupt.edu.cn) is the corresponding author.}
}

\maketitle

\begin{abstract}
The prevalent engagement with mobile apps underscores the importance of understanding their data practices. Transparency plays a crucial role in this context, ensuring users to be informed and give consent before any data access occurs. Apple introduced a new feature since iOS 15.2, App Privacy Report, to inform users about detailed insights into apps' data access and sharing. This feature continues Apple's trend of privacy-focused innovations (following Privacy Nutrition Labels), and has been marketed as a big step forward in user privacy.
However, its real-world impacts on user privacy and control remain unexamined. We thus proposed an end-to-end study involving systematic assessment of the App Privacy Report's real-world benefits and limitations, LLM-enabled and multi-technique synthesized enhancements, and comprehensive evaluation from both system and user perspectives. Through a structured focus group study with twelve everyday iOS users, we explored their experiences, understanding, and perceptions of the feature, suggesting its limited practical impact resulting from missing important details. We identified two primary user concerns: the clarity of data access purpose and domain description. In response, we proposed enhancements including a purpose inference framework and domain clarification pipeline. We demonstrated the effectiveness and benefits of such enhancements for mobile app users. This work provides practical insights that could help enhance user privacy transparency and discusses areas for future research. 
\end{abstract}


\section{Introduction}
Mobile applications (apps) have become an essential part of our daily lives. From networking and entertainment to productivity and business, they play a vital role in how we interact with technology. However, the increasing reliance on mobile apps raises privacy concerns, making understanding apps' data access practices and the potential risks become paramount \cite{mediumreport}.
Apps typically engage in accessing, storing, and processing various forms of personal and sensitive data such as personal identifiers, geographic locations, and contact lists~\cite{datacollection}. 
Without clear understanding of app purposes and behaviors, many users blindly grant permissions to apps' data access without knowledge of the consequences, exposing themselves to data exploitation~\cite{polykalas2019mobile}.

To address such concerns, many authorities have enacted laws and regulations to protect user privacy. Notable examples include EU's General Data Protection Regulation (GDPR)~\cite{GDPR}, the California Consumer Privacy Act (CCPA)~\cite{CCPA} in the US, and China's Personal Information Protection Law (PIPL)~\cite{PIPL}. 
In recent years, app stores, as gatekeepers of app distribution, have also invested considerable effort in enhancing user privacy and enforcing transparent app data practices.
For example, the Apple App Store introduced mandatory ``\textit{Privacy Nutrition Labels}''~\cite{privacylabel} for iOS apps in December 2020, requiring developers to disclose what data their apps collect and for what purpose.
Similarly, Google Play introduced the ``\textit{Data Safety}''~\cite{datasafety} section for Android apps in July 2022.
Despite these initiatives, there has been growing evidence showing the limitations of simple privacy labels, for example, inconsistencies between privacy labels and actual privacy/data practices~\cite{khandelwal2023overview,xiao2023lalaine,koch2022keeping}. This indicates that some app developers may not fully self-disclose their actual data practices.
Moreover, some researchers found that these labels are not as usable as they were supposed to be, with many users struggling to comprehend the information on privacy labels~\cite{zhang2022usable,lin2023data}. 

\begin{figure*}[htbp]
    \centering
    \includegraphics[width=0.98\textwidth]{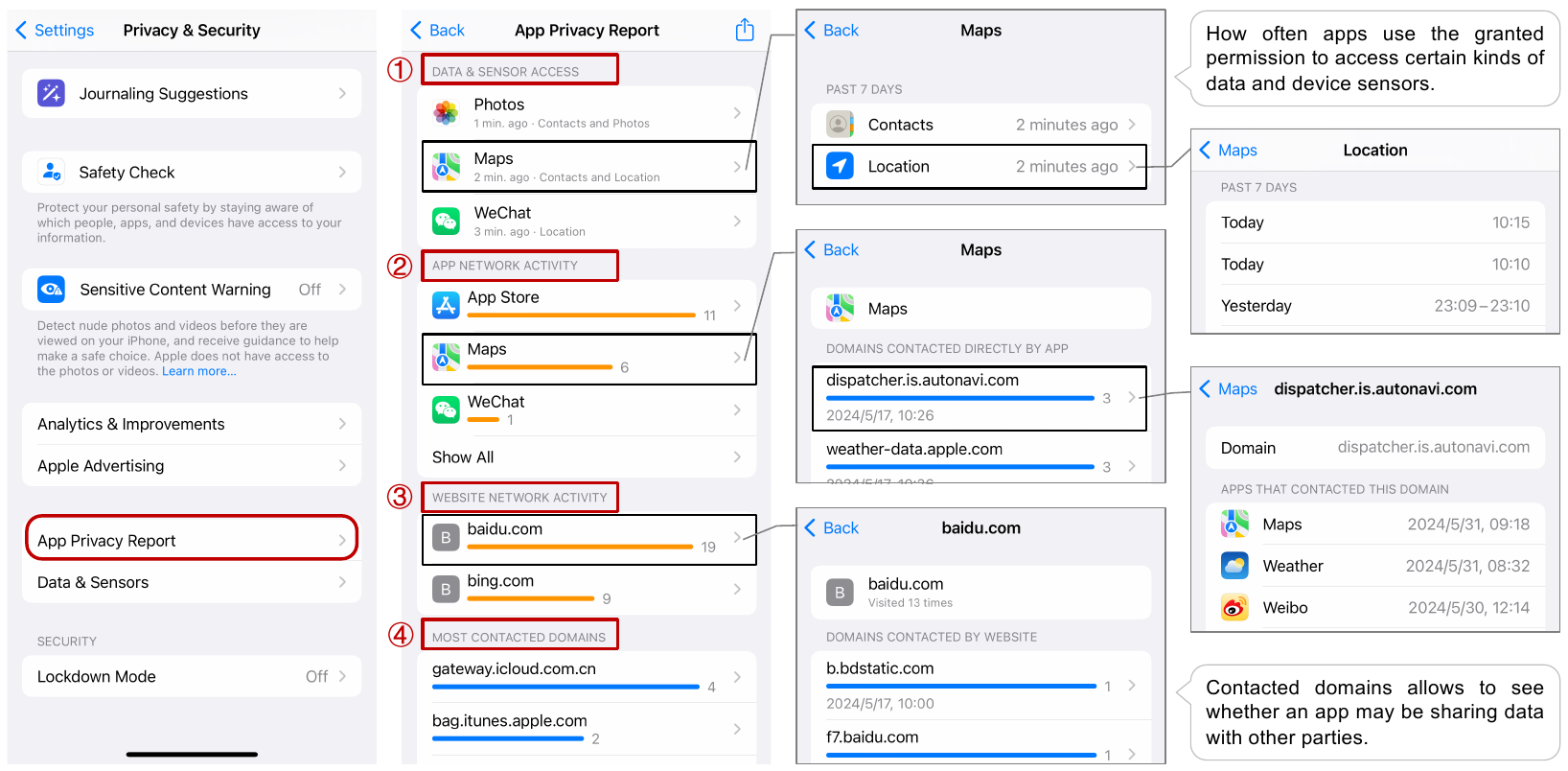}
    \vspace{-0.1in}
    \caption{An overview of the App Privacy Report feature on an iPhone running iOS 17.6.1.}
    \vspace{-0.1in}
    \label{fig:report}
\end{figure*}

In 2021, Apple introduced a new ``App Privacy Report'' feature from iOS 15.2~\cite{APR}, designed to offer more visibility of what all the apps are really up to in the background of the device. 
As shown in \autoref{fig:report}, this new page is nestled in \textit{Settings} $\rightarrow$ \textit{Privacy} $\rightarrow$ \textit{App Privacy Report}. 
As advertised, the App Privacy Report allows users to see what private data each app is accessing on the device and how often it is happening, as well as what domains or website addresses are being accessed (detailed in \S~\ref{sec:privacyreport}).
It goes beyond the potentially fallible privacy labels and instead works to collect information about how apps are behaving more directly. 
According to \textit{Erik Neuenschwander}, Apple's director of user privacy, ``\textit{App Privacy Report gives users new insights into the behavior of apps on their device, and is a good complement to apps’ Privacy Nutrition Labels. It helps users make well-informed choices about the apps they use, how they use them and their privacy settings}''~\cite{wpost}.

However, such a feature's real-world impacts remain unclear, with potential limitations yet to be fully understood.
So far, researchers and practitioners know little about whether and to what extent the App Privacy Report fulfills its promises as it is designed to be.
We thus conducted an end-to-end study to explore the feature's real-world impacts, reveal its shortcomings, design novel enhancements to address these shortcomings, and evaluate the enhancements with real-world users.
We first organized a structured focus group meeting with 12 iOS users to explore their experiences, perceptions, attitudes, and expectations concerning the App Privacy Report (see \S~\ref{sec:focusgroup}). The focus group meeting identified shortcomings in the App Privacy Report that hindered the feature's usefulness, particularly regarding the clarity of the purpose of data access and the description of domain names. 
Motivated by this user feedback and the transparency requirements mandated by privacy regulations~\cite{GDPR,CCPA,PIPL}, we propose several enhancements to the feature (see \S~\ref{sec:approach}). 
First, we propose a lightweight framework that can automatically infer the purpose of data access during runtime.
The framework encodes two levels of heterogeneous sources as input, i.e., dynamic code behavior (code-level), and app description and privacy policies (text-level), leveraging large language models (LLMs) to reason about purposes.
Second, we introduce a allowlist-based method to continuously collect domains and their descriptions for improved user awareness. 
Our evaluations, which include performance and user experience testing, demonstrate the feasibility and utility of these enhancements (see \S~\ref{sec:evaluation}).
Our work offers design enhancements that could be adopted by mobile systems and highlights areas for future research. The data and code of our study are available at~\cite{repo}.
In summary, this paper makes the following contributions:

\begin{itemize}
    
    \item We thoroughly examine the ``App Privacy Report'' feature in iOS devices from a user-centric perspective. Based on an in-depth focus group study, we reveal that the current App Privacy Report offers limited assistance in helping users manage privacy. We identified two main user concerns that guided the design of an enhanced version of the feature.

    \item We propose promising enhancements in response to user concerns. Firstly, we combine dynamic analysis and prompt engineering to harness LLMs' capabilities for automated inference of data access purposes. Secondly, we develop a semi-automated pipeline that leverages both external and LLMs' knowledge to create an evolving domain allowlist for users.

    \item We conduct experiments to evaluate the performance of proposed enhancements.
    The results, based on real user interactions with iPhones, confirm the feasibility and effectiveness of our methods. Additionally, a Think Aloud study further demonstrates the significant benefits of suggested enhancements.
\end{itemize}

Beyond these specific contributions, our paper provides a structured approach for systematically evaluating the usability and utility of new privacy engineering features. We demonstrate a user-centered approach--from gathering feedback to testing enhancements--and show how LLMs can be used to bridge gaps between functionality and user-friendliness.
This framework offers a referable model and workflow for researchers and practitioners seeking to evaluate and enhance usability of privacy engineering features in diverse digital contexts.

\section{Background \& Motivation}
\label{sec:background}

\subsection{Need for Data Transparency in Mobile Apps}
In our context, data transparency refers to clearly and openly informing users about how their data is collected, used, and shared by mobile apps. This is critical for building user trust and enabling informed decision-making.
Researchers have highlighted a significant gap between user perceptions and the actual privacy risks associated with app usage.
For example, studies~\cite{felt2012android,chin2012measuring} showed that mobile users often have a poor understanding of permissions, e.g., unable to correctly interpret the implications of granting certain permissions.
Lin et al.~\cite{lin2012expectation,lin2014modeling} revealed a semantic gap between users' expectations and app behaviors, indicating that users' mental models of how apps should behave do not align with the apps' real data practices. 
Therefore, there is a critical need to bridge the gap between user awareness and app behaviors through transparency on data practices. 
When users are informed about the specific behaviors of apps accessing certain sensitive resources, they are better equipped to make informed decisions about their privacy.

\subsection{Advancements and Challenges of Privacy Features in Mobile Platforms}

Mobile operating systems (e.g., iOS and Android) have progressively introduced features aimed at enhancing data transparency and user control over personal information. 
One of the earliest and most fundamental features is the privacy policy, in which apps are required to provide users with detailed information about their data practices in a detailed and thorough document. Over time, mobile platforms have progressively expanded their privacy tools to offer users greater visibility and control over how their data is used.
Apple, in particular, has introduced a series of important privacy features, such as \textit{Privacy Nutrition Labels}~\cite{privacylabel} (from iOS 14), which provide a summary of an app's data practices displayed on the App Store; 
\textit{Safari Privacy Report}~\cite{Safari-pr} (from iOS 14), which shows users the trackers blocked by Safari’s Intelligent Tracking Prevention (ITP); 
\textit{App Tracking Transparency}~\cite{att} (from iOS 14.5), which requires apps to request user consent before tracking their activity across other apps and websites; and \textit{Mail Privacy Protection}~\cite{mail} (from iOS 15), which blocks tracking pixels and hides users' IP addresses to prevent email senders from profiling and tracking their location. \textit{These features reflect Apple's ongoing efforts to improve user privacy by making data practices more transparent and empowering users to make informed decisions about their data.}

While these initiatives represent a significant step forward in promoting transparency, they still face challenges in terms of usability, accessibility, and user engagement.
Studies found there is a risk that many users just ignore the new privacy nutrition labels as they commonly do with privacy policies~\cite{kollnig2022goodbye} and some developers may not honestly self-declare their actual data practices in privacy labels~\cite{lin2023data}. 
Similarly, App Tracking Transparency (ATT) prompts can be misleading, with some apps using deceptive strategies to persuade users to grant tracking permissions~\cite{mohamed2024attention}.
Meanwhile, researchers have dedicated considerable effort to improving the usability of those privacy notices. For example, a significant body of research has focused on privacy policies, including enhancing the automatic annotation and extraction of specific information~\cite{sathyendra2017identifying}, assessing the quality and transparency of privacy policies~\cite{liao2020measuring}, and automating compliance analysis~\cite{xie2022scrutinizing}. For privacy nutrition labels, researchers have explored methods to design clearer and more intuitive presentations~\cite{zhang2024exploring}, develop automated generation techniques~\cite{pan2023toward}, and analyze (in)consistencies across privacy disclosures~\cite{khandelwal2023overview,xiao2023lalaine,koch2022keeping}.
\textit{These efforts underscore the need for continuous refinement of privacy transparency tools to ensure they are not only informative but also accessible, trustworthy, and effective in helping users make privacy-conscious decisions.}

\subsection{iOS App Privacy Report}
\label{sec:privacyreport}
With the release of iOS 15.2, Apple introduced the ``App Privacy Report'' feature to further enhance data transparency.
This feature is ``off'' by default and users need to turn it on and grant the necessary permissions for data collection to generate the report. 
Once enabled, it tracks the activities of individual apps and websites immediately and delivers the report in real time\footnote{Although it takes time for the phone to collect the data and compile the report, we can usually see results within seconds.}. It records up to seven days of activity.
The Privacy Report consists of four sections:
\ding{172} \textbf{Data \& Sensor Access} shows a summary of how often and when an app has accessed sensitive information like contacts, camera, location, microphone, photos, and media library, along with timestamped logs. The list is in reverse chronological order, so the most recent one is at the top.
\ding{173} \textbf{App Network Activity} shows the list of web domains that each app has contacted. It shows both internal and third-party domains used by apps, allowing users to figure out what third-party services/websites (e.g., analytics and tracking tools) are contacted in the background.
\ding{174} \textbf{Website Network Activity} displays a similar list but focuses on websites that contacted domains, some of which may have been provided by an app. It includes all the domains that user-visited webpages contacted, whether the user viewed them in browsers (e.g., Safari, Google Chrome) or other apps. 
\ding{175} \textbf{Most Contacted Domains} shows a list of domains ranked by the number of different sites and apps that contacted them, highlighting frequently accessed domains. 
Broadly, the last three sections are very similar in that they tell you what domains or website addresses are being accessed by the device.

In summary, the App Privacy Report represents a significant step towards data transparency that provides users with comprehensive insights into data access and network activity over the past seven days, ensuring they are fully informed about their app's access to sensitive data and where data is shared. This functionality is undoubtedly beneficial and well-intentioned, but exactly how well it works in real-world scenarios remains largely unexplored.
This study is among the first attempts seeking to fill this research gap by examining the App Privacy Report from the perspective of everyday iOS users and providing end-to-end design enhancements to tackle its potential limitations.

\section{Focus Group Study}
\label{sec:focusgroup}

\begin{figure}[htbp]
    \centering
    \includegraphics[width=0.48\textwidth]{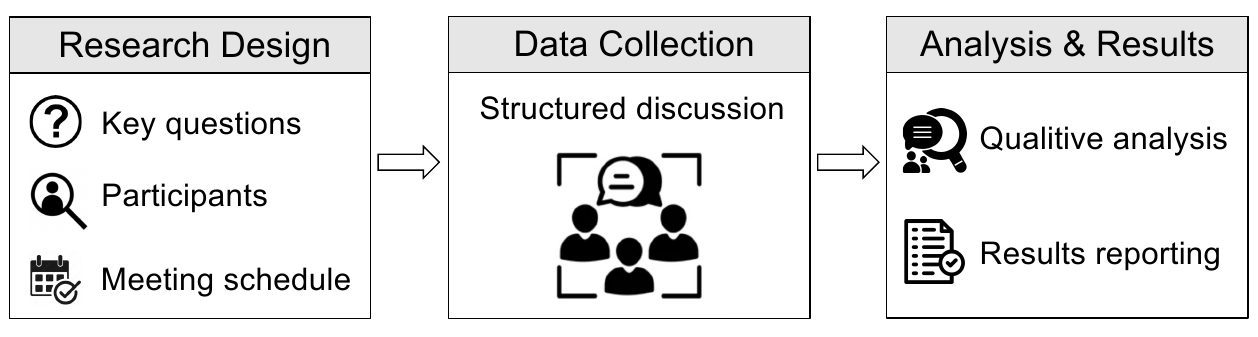}
    \vspace{-0.2in}
    \caption{An overview of the focus group study design.}
     \vspace{-0.1in}
    \label{fig:focusgroup}
\end{figure}

To understand its practical effectiveness, we first organized a structured focus group meeting to discuss the topic around the iOS App Privacy Report.
The focus group meeting method was chosen for its efficiency in data collection and its ability to uncover participants' collective understanding of the issues, making it particularly valuable for investigating subjective meanings and perspectives~\cite{barnes2012does}. Additionally, this method facilitates the generation of ideas and reflections that participants might not have considered individually, making it well-suited for our topic.
Our focus group study took place in April 2024. 
Following related literature~\cite{o2018use}, our study was designed into four phases: (1) research design, (2) data collection, (3) analysis, and (4) reporting of results, as per \autoref{fig:focusgroup}.

\subsection{Research Design}

\subsubsection{Define the Objectives of the Study}
The process begins with defining the key research objectives of the focus group study, which is to understand real users' feelings and opinions about the iOS App Privacy Report.  
To this end, a list of questions (schedule) is prepared as guidance for the session. 
As shown in \autoref{tab:questions}, we outlined four main topics with 16 specific questions designed to elicit participants' feedback.
It began with examining participants' sense of privacy to gain an initial understanding of their privacy preferences.
We then explored their awareness and usage of the Privacy Report, followed by their experiences and perceptions of the feature. 
Finally, we discussed their attitudes and expectations regarding the content in the Privacy Report, including their level of trust, perceived necessity, and suggestions. 
This structured, drill-down approach provided comprehensive insights into users' privacy-related perspectives.

\subsubsection{Ethical Considerations}
Potential ethical risks have been well considered in our research. Participants were given informed consent, their data was anonymized, and recordings were securely stored. 
They could withdraw at any time without consequences.
This study has been reviewed and approved by the Academic Ethics Review Committee of the first author's institution whose role is equivalent to the Institutional Review Board (IRB) in the United States. 

\begin{table*}[t]
    \centering
    \caption{Topics and questions for the focus group discussion.}
    \vspace{-0.1in}
    \resizebox{1.0\textwidth}{!}{
    \begin{tabular}{c|l}
    \toprule
    \textbf{Topic} & \textbf{Questions for participants to discuss} \\ \midrule
    \multirow{3}{*}{\shortstack{Sense \\of \\Privacy}}  & 1. How important is privacy to you when using apps on your iOS device? \\
      & 2. On a scale from 1 to 10, how would you rate your concern about app privacy? \\
      & 3. What steps do you usually take to protect your privacy on your phone? (e.g., changing settings, reviewing app permissions) \\ \midrule
    \multirow{4}{*}{\shortstack{Awareness \\and \\Usage}} & 4. Have you heard about the App Privacy Report feature? If so, how did you first learn about it? \\
      & 5. Have you enabled the App Privacy Report on your devices? If yes, what motivated you to enable it? \\
      & 6. How often do you check your App Privacy Report? \\
      & 7. Have you made any changes to your app usage or privacy settings based on the information provided in the App Privacy Report? \\ \midrule
    \multirow{5}{*}{\shortstack{Experiences \\and \\Perceptions}} & 8. Can you fully understand the contents provided in the App Privacy Report? \\
      & 9. How do you feel about the level of detail provided in the App Privacy Report? \\
      & 10. What specific information or section do you find most useful in the App Privacy Report? \\
      & 11. Do you think the App Privacy Report has made you more aware of privacy issues related to the app? Why or why not? \\
      & 12. How do you perceive the importance of such a feature in terms of your overall app usage and privacy? \\ \midrule
    \multirow{4}{*}{\shortstack{Attitudes \\and \\Expectations}} & 13. Do you trust the accuracy and completeness of the information provided in the App Privacy Report? Why or why not? \\
      & 14. What are your general thoughts about the needs for features like the App Privacy Report? \\
      & 15. What kind of support or information would help you better understand and utilize the App Privacy Report? \\
      & 16. What additional features or enhancements would you like to see in the App Privacy Report? \\
      \bottomrule
    \end{tabular}}
    \label{tab:questions}
\end{table*}

\subsubsection{Identify and Recruit Participants}
Participants were recruited from a mid-size public university in China. Eligibility criteria required participants to be proficient in using mobile apps and to own an iOS device.
The study was advertised on campus through various channels (e.g., social media and campus forums). 
As for the size of the group, research~\cite{krueger2014focus} suggested that 10 participants is generally considered large enough to gain a variety of perspectives and small enough not to become disorderly or fragmented. To account for potential no-shows, we overly-recruited by 20\% (12 participants).
Participants were contacted individually via SMS to schedule the meeting.
All 12 recruited participants attended the focus group session.
The group comprised 6 males and 6 females, with 7 undergraduates and 5 postgraduates, aged between 20 and 25. Each participant possesses at least one iOS device (iPhone with iOS version $\textgreater$ 15.2), and all had been using iOS for more than one year, with two-thirds having more than two years of experience and one-third having more than three years.
Although from the same university, the participants have diverse professional and academic backgrounds, balanced gender representation, and varying levels of iOS device experience. 
Thus, the sample has reasonably good representativeness within the demographic of young, educated users.

\subsection{Data Collection}

\subsubsection{Pre-session Preparation}
Participants were asked to bring their own iOS devices to the meeting to facilitate necessary interactions with the devices during the session.
Upon arrival at the meeting room, participants were greeted by the research team and provided with a brief overview of the discussion process.
We also requested permission from each participant to video and audio record them and asked them to review and sign a consent form.
Participants were allowed to ask us the researchers any questions before proceeding to ensure they fully understood the tasks ahead.

\subsubsection{Facilitation During Meeting}

This phase was crucial for exploring participants' experiences, perceptions, and understanding of the App Privacy Report. 
The meeting was moderated by a senior researcher (moderator) accompanied by a PhD candidate (assistant). 
The moderator guided the discussion through predefined questions, allowing participants to elaborate on themes directly related to the topics but brought the discussion back to the topic when there was a risk of losing the lead.
The assistant observed verbal and nonverbal cues without participating directly.
The meeting began with a welcoming statement and introductions, then discussed centering on four main topics, and ended with an open invitation for any final comments or suggestions, ensuring that all relevant insights were captured. 

\noindent \textbf{(1) Study Briefing and Warm-Up.}
The moderator first greeted the participants, expressed gratitude for their participation, and introduced the session's purpose. Then a brief ice-breaker activity was conducted to ease participants into the discussion, where participants introduced themselves and shared one aspect of their daily routine that involves their iOS devices.
This initial step was essential for setting a positive tone and establishing rapport with the participants.

\noindent \textbf{(2) Structured Group Discussion.}
The discussion followed the predetermined series of questions, progressing from surface-level inquiries to more profound explorations.
The first segment focused on the participants' sense of privacy (Q1-3), exploring their general attitudes toward mobile privacy. 
The remaining segments were all centered around the App Privacy Report feature. 
We began with awareness and usage of the App Privacy Report (Q4-7), assessing the participants' awareness and familiarity with the feature. 
The next segment transitioned to participants' experiences and perceptions of the App Privacy Report (Q8-12). As these questions presupposed that participants were aware of the App Privacy Report, before proceeding with the questions, participants were given time to interact with their devices, particularly those who were unfamiliar with the feature. 
They were recommended to follow these steps to interact:
1) Participants ensured that the App Privacy Report had been enabled, and selected several apps they frequently use for interaction. 
2) Participants launched an app and interacted with it as usual, handling permission requests as they normally would.
3) After playing for a while, they checked the App Privacy Report and tried to understand the reported results. They repeated this process to gain a better understanding. 
There was no time limit, and they were free to work as long as they liked until they felt ready for the follow-up discussion.
As a result, they basically completed their interactions within six minutes. The discussion then continued, with participants sharing their feelings about the level of detail provided in the Privacy Report, their ability to understand its contents, the usefulness of specific information, etc.
The final segment delved into participants' attitudes towards the App Privacy Report and their expectations for future enhancements (Q13-16). This topic covered participants' trust in the Report's content, general views on the necessity of such privacy-related features, support needed for using the Privacy Report, and suggestions for additional features or enhancements. The last two questions were at the forefront of our concerns, and we gave participants ample time for reflection and sharing views.

\noindent \textbf{(3) Conclusion.}
The session concluded with an open invitation for any final comments or suggestions regarding the App Privacy Report feature. The moderator summarized the key points of the discussion and thanked the participants for their valuable feedback and insights.
The entire discussion process took 70 minutes. Before leaving the meeting, each participant received a flat rate of 70 Chinese Yuan ($\approx$ 9.66 USD) as compensation.

\begin{figure}[htbp]
    \centering
    \includegraphics[width=0.46\textwidth]{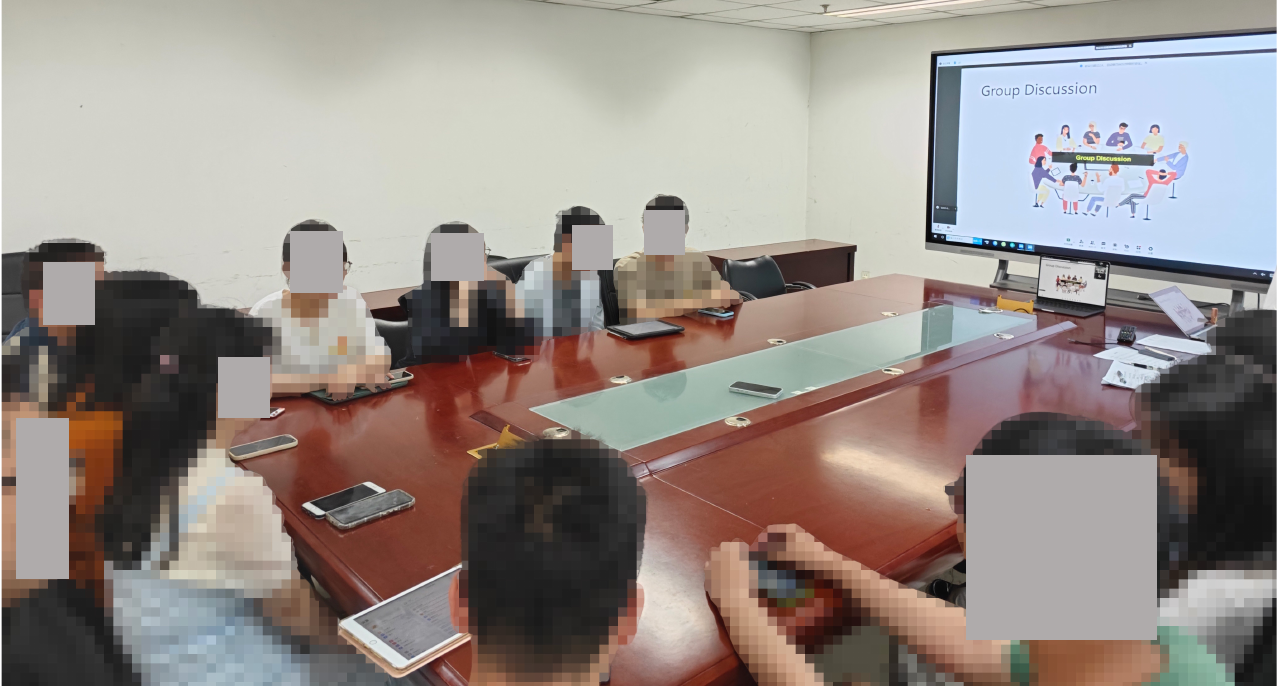}
    \vspace{-0.05in}
    \caption{Participants (identities obscured) were at the focus group meeting.}
     \vspace{-0.1in}
    \label{fig:domain}
\end{figure}

\subsection{Analysis}

The entire group discussion session was videotaped, and participants' comments were transcribed. We then performed a qualitative analysis of the transcribed text through an inductive data coding process. We followed the method in Charmaz~\cite{charmaz2006constructing} to conduct detailed coding of participants' words. The first step, initial coding, involved generating numerous category codes without limiting their number.
We started without a predefined codebook; instead, initial codes and themes were derived directly from participants' comments.
For each question, the researcher identified participants' attitudes, listed their ideas, and marked frequently used keywords as indicators of important themes. The second stage, focused coding, involved refining the initial codes by eliminating, combining, or subdividing them. Attention was given to recurring ideas and broader themes that connected the codes.
The final codebook was developed during this phase.
The data coding process was independently conducted by the first two authors of this study. Any disagreements were resolved through a meeting with all authors. Finally, different participants' opinions were interconnected and compared to draw insights.

\subsection{Results and Findings}

\vspace{-0.05in}
We next report the results of the focus group discussion around the App Privacy Report.

\vspace{-0.05in}
\subsubsection{Participants' Sense of Privacy}
All participants emphasized the importance of personal privacy in mobile apps, especially concerning personally identifiable information and bank card details. In response to Q2, 10 out of 12 participants rated their privacy concerns higher than 6 on a 10 scale (with the remaining 2 participants rating them as 5). The mean, median, and standard deviation of participants' privacy concern ratings are 6.83, 7, and 1.34, respectively\footnote{These statistics are reported to provide a general intuition about participants' attitudes; however, given the small sample size, they are not intended to support any statistically significant conclusions.}. Half of the participants shared measures they took to protect their privacy, such as altering permission settings, modifying system location data, and restricting access to photos, indicating a general sense of privacy among them.

\subsubsection{Awareness and Usage}
None of the participants had previously heard of the App Privacy Report feature on iOS. Before the meeting, only one participant (P5) had this feature enabled on his iPhone but quite without intention. Apparently, they had never checked the Privacy Report before, let alone made any changes to their privacy settings based on the information provided.
This was unexpected since most participants claimed to be concerned about user privacy, and some of them were Apple fans, who should be familiar with Apple's various features.
The ensuing discussion may reveal some possible reasons: some participants mentioned that they rarely paid attention to what exact features were updated so they missed this new feature; some expressed a general trust in Apple's privacy protections making them not specifically seek out privacy-related features on their phones.
Anyway, this suggests that users' actual benefits of this feature may be much lower than expected.
It underscores the need for proactive user education and clear communication about new privacy tools.

\subsubsection{Experiences, Perceptions, Attitudes, and Expections}
After interacting with the apps and App Privacy Report, all participants agreed that this feature is helpful with data transparency. However, they also expressed some concerns about it, and suggested potential design spaces for enhancements. 

\noindent \textbf{Difficulty in Interpreting Some Contents.}
For Q8, most participants indicated that they could not fully understand the contents in the Privacy Report. Especially for the domains (listed under ``App/Website Network Activity'' and ``Most Connected Domains''), which require users to have certain technical expertise to identify and interpret. For example, a participant remarked: ``\textit{There are just so many obscure domain names. It is very difficult for normal users to know what these domains and IPs are related to and whether their connections are necessary.}'' [P6]. Thus the last three network parts might not be that user-friendly.

\noindent \textbf{Desire for More Details.}
Many participants were not satisfied with the level of detail provided in the Privacy Report (Q9). As mentioned above, they suggested that it would benefit end users if Apple added some additional context about domains, such as the owner, so that users know who (which company or service) the app is accessing. Besides, the participants paid attention to the ``Data \& Sensor Access'' section, reflecting that while this part enables users to see how often each app accesses sensitive data but fails to specify what this data is used for, i.e., the purpose. 
There was a consensus on the need for clarity on why certain data is required, especially when the purpose is not obvious. For instance, a participant noted,
``\textit{Sometimes the purpose is guessable--like a mapping app using location for navigation. But why would a flashlight app need my location?}'' [P4].

\noindent \textbf{Varied Opinions on Most Useful Sections.} For Q10, half of the participants considered the first section most important because it allows them to manage settings if any data/permission misuse is found. The other half prioritized the network activity sections. Despite limited knowledge of domain names, they believed that network activity offers significant insights into the app's behavior behind the scenes.
This suggests that users prioritize different aspects of the feature, demonstrating the value of its multifaceted design--no part is considered dispensable.

\noindent \textbf{Positive Attitudes Toward the Feature.}
For Q11 and Q12, all participants agreed that the Privacy Report could raise their awareness of privacy issues related to apps and recognized the importance of such a feature for users' overall app usage. 
For Q13, participants generally trusted the accuracy and completeness of the information (data accesses and domain connections) provided in the Report as it comes from official surveillance.
Responses to Q14 indicated a positive attitude toward the necessity of such features for everyday use.
Moreover, all of them reported that they would use and review the Privacy Report in the future.
These findings indicate strong adoption potential for the feature, and further underscore the importance of user-centered design and usability.

\noindent \textbf{Suggested Enhancements.}
Despite recognizing its importance, participants felt the App Privacy Report was not as usable or effective as it could be in its current form and configuration. 
They proposed some ideas for enhancements. A key recommendation, as discussed earlier, was to specify the rationale and purpose behind accessing each sensitive data in the ``Data \& Sensor Access'' section.
Moreover, participants expressed difficulty in understanding many of the domain contacts and suggested the Report should indicate the type and service of the accessed domain or site. 
These two suggestions were the most widely endorsed among participants.
In addition, some participants also offered other suggestions. For example, several participants recommended refining the granularity of the data presented and broadening the scope of the monitoring information covered in the current Report (see \S\ref{sec:concerns}). 
These suggestions provide valuable insights for enhancing the feature to better inform and empower users.

\begin{tcolorbox}[title=Major Insights from the Focus Group Study, left=2pt, right=2pt, top=2pt, bottom=2pt]
    The focus group discussion allows us to understand, for the first time, how everyday iOS users perceive and engage with the iOS App Privacy Report.
    Although the App Privacy Report is considered a major step forward in privacy transparency, our focus group study revealed several obstacles that limit its effectiveness in truly assisting users. Particularly, the key issues identified include:
    \begin{itemize}
        \item Usability Limitation \#1: the need for clearer explanations of data usage purposes.
        \item Usability Limitation \#2: the need for specific information regarding domain contacts.
    \end{itemize} 
    Encouragingly, all participants acknowledged the importance of this feature and expressed a willingness to use it. Therefore, this feature has great potential, and the company should consider user feedback to make necessary enhancements in the future.
\end{tcolorbox}

\section{Enhancements}
\label{sec:approach}

The focus group study identified several key limitations of the current Privacy Report from the perspective of real-world users.
Most notably, users expressed two aforementioned usability limitations: the lack of clear explanations regarding the purposes of data usage and the intentions behind domain contacts. 
This communication gap is particularly concerning, as it limits users' ability to fully understand how their data is used--an essential aspect of transparency emphasized by
privacy regulations, such as GDPR Article 13~\cite{GDPR}, CCPA §1798.130(a)(5)~\cite{CCPA} and PIPL Article 17~\cite{PIPL}.
Hence, these observations motivate us to develop approaches to enhance the user-concerned details of the current App Privacy Report. To address Limitation \#1, we propose inferring and articulating the purpose of an app's sensitive data usage at runtime (\S~\ref{sec:pur-inf}). For Limitation \#2, we propose providing clearer information about the types and services of the accessed domains and websites (\S~\ref{sec:dom-cla}).
By tackling these issues, we aim to significantly enhance the utility and compliance of the App Privacy Report, ensuring that users are better informed and more empowered to manage their privacy.

\subsection{Purpose Inference}
\label{sec:pur-inf}
Determining the purpose of data access at runtime is a non-trivial task due to the dynamic and context-dependent nature of mobile app behavior.
Below, we describe the key challenges and our insights in addressing these challenges.

\noindent \textbf{Challenges.}
To be specific, there are three key challenges in achieving automated runtime purpose inference:

\begin{itemize}[noitemsep, topsep=1pt, partopsep=1pt, listparindent=\parindent, leftmargin=*]

 \item[\ding{172}] \textit{Runtime Permission Monitoring.} 
 Continuously tracking how and when sensitive data is used by an app at runtime is challenging. Data might be accessed dynamically, conditionally, or in response to specific user interactions, making static analysis insufficient.
  
 \item[\ding{173}] \textit{Diverse Data Controllers.} 
  Sensitive data can be accessed by various entities, including the app itself (first party) or a variety of third parties, serving different purposes within an app. For example, location data might be used for navigation, targeted advertising, or social networking. Distinguishing these purposes requires understanding who (first or third party) is using the data.

  \item[\ding{174}] \textit{Functionality and Context Dependency.}
  Inferring the purpose of data access necessitates understanding the app's functionality and context. The data can be used for different purposes depending on the app. For instance, a navigation app might access location for route planning, while a weather app might use the same data to provide local weather updates.

\end{itemize}

\noindent \textbf{Key Design Rationale.}
To address these challenges, the approach is powered by three design rationales:

\begin{itemize}[noitemsep, topsep=1pt, partopsep=1pt, listparindent=\parindent, leftmargin=*]
 \item \textit{Dynamic Instrumentation of System APIs.} 
 In the iOS environment, apps must use designated APIs to request permissions for accessing sensitive user data like photos, contacts, and location. These APIs grant access to such data during runtime, contingent upon explicit user consent. 
 As such, dynamic instrumentation of these APIs enables runtime monitoring of sensitive data usage. (C\ding{172})

 \item \textit{Analysis of API Call stack trace.}
 A stack trace shows the chain of function or method calls that led to a specific API invocation. Analyzing stack traces can uncover the code paths and conditions under which the API is invoked. The naming conventions of the classes and methods in the stack trace can help identify the data controller and specific functionality. 
 (C\ding{173}, C\ding{174})

 \item \textit{Integration of App Description and Privacy Policy.} 
 The natural language description of an app serves as a proxy for its advertised features~\cite{gorla2014checking}. 
 Moreover, privacy policies contain details such as what data are collected, what permissions are required, and what the data are used for. Both app descriptions and privacy policies are valuable resources providing context about data usage. (C\ding{174})

\end{itemize}

Based on these key insights, we decided to encode three heterogeneous sources as input, i.e., the sensitive API call stack traces, app description, and privacy policy, extracting and integrating their features to infer the possible purpose of data usage. We next elaborate on our approach.

\subsubsection{Architecture Overview}

\begin{figure}[htbp]
    \centering
    \includegraphics[width=0.48\textwidth]{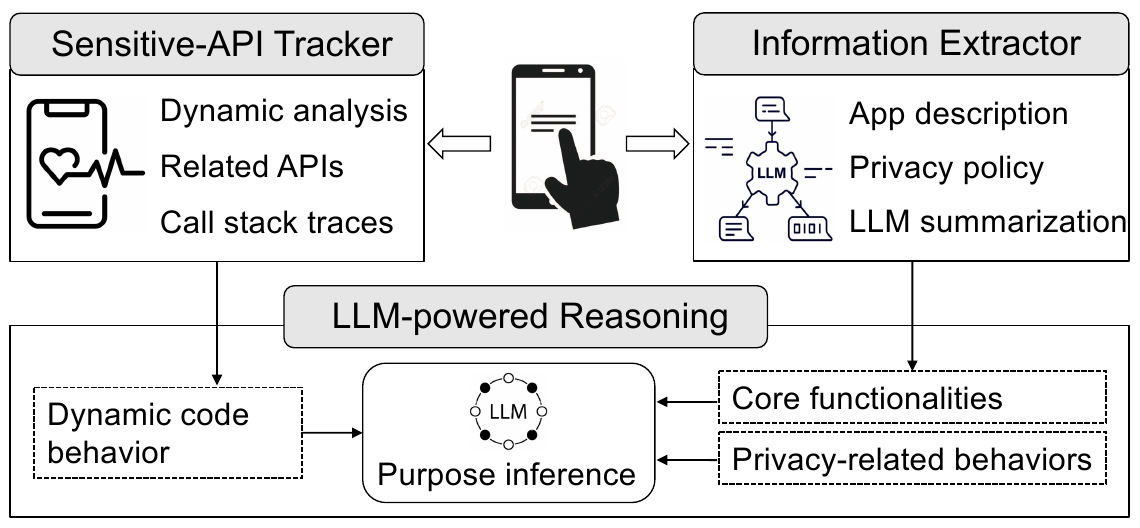}
    \vspace{-0.2in}
    \caption{Architecture of the purpose inference framework.}
    \label{fig:purpose}
\end{figure}

The overview of our approach is shown in \autoref{fig:purpose}.
It combines the features of dynamic code behavior, app descriptions, and privacy policies to infer the purpose of data usage. 
The underlying method to achieve this combination is LLMs and prompt engineering, i.e., prompting the LLM to generate the desired output, which is very lightweight. 
In particular, the framework includes three major components: 
1) \textit{sensitive-API tracker}, which performs dynamic instrumentation of systems APIs and encodes code-level semantic features to infer possible purposes of data usage; 2) \textit{information extractor}, which extracts relevant information from documents and encodes text-level features of the app's functionality and privacy behaviors to enhance purpose explanations; 3) \textit{LLM-powered reasoning}, which sequentially provides LLM with knowledge of code-level and text-level features, prompting it to reason about the most likely purpose and give a summary. 
We next describe each component in detail.

\begin{figure}[htbp]
    \centering
    \includegraphics[width=0.48\textwidth]{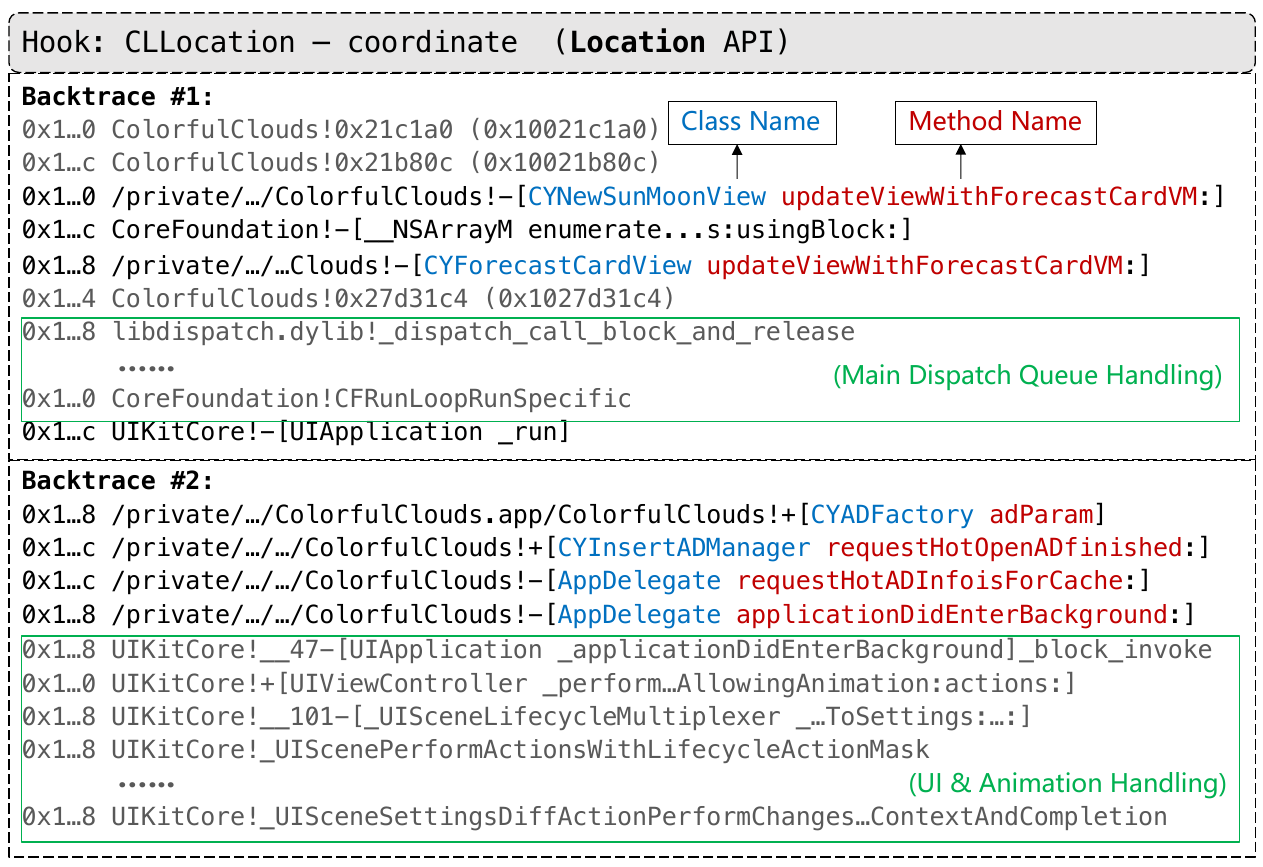}
    \vspace{-0.2in}
    \caption{Two examples of a location API's call stack traces.}
    \vspace{-0.1in}
    \label{fig:stacktrace}
\end{figure}

\subsubsection{Sensitive-API Tracker}
As aforementioned, access to sensitive data is triggered by permission-protected API (or sensitive API) calls, and call stack traces can help identify the actual data controller (first party or third party) and specific functionality, thereby inferring the intent behind the data access. 
For example, \autoref{fig:stacktrace} illustrates two invocations of the location API \texttt{CLLocation - coordinate} in the ColorClouds Weather app. The stack traces detail the sequence of method calls in each situation.
In the first trace, we can see that the call originates from the \texttt{CYNewSunMoonView: updateViewWithForecastCardVM} method, suggesting that location data may be used to update weather forecast views. 
In the second stack trace, the call originates from \texttt{CYADFactory: adParam}, and is followed by a series of calls including \texttt{CYInsertADManager: requestHotOpenADfinished}, implying that the app is preparing and requesting advertisements, potentially personalized based on the user's location. 
This differentiation underscores the importance of understanding the controller and function behind data access.

Thus, we implement a sensitive-API tracker to dynamically track the invocation of sensitive APIs and their call stack traces.
Specifically, we leverage Frida~\cite{frida}, a dynamic instrumentation toolkit, to hook into iOS systems APIs at runtime and retrieve call stack traces using the \texttt{Thread.backtrace([context, backtracer])} method. 
This monitoring focuses on APIs related to location, camera, microphone, contacts, photo library, media library, and screen recording -- key permissions highlighted in the App Privacy Report. 
The tracked information is logged, and to minimize redundancy, we eliminate duplicated call stacks within a specified time window. The refined call stack data is then forwarded to the LLM-powered purpose reasoning component, which further analyzes the call stack by integrating it with pre-processed data from app descriptions and privacy policies (see \S~\ref{sec:info-extractor}) to deduce the purpose behind the used permissions.

\subsubsection{Information Extractor}
\label{sec:info-extractor}
As aforementioned, app descriptions and privacy policies are useful resources for our purpose inference task. 
However, they often contain an overload of information for our specific needs, necessitating the extraction of only the relevant details. Specifically,
in app descriptions, we focus on functionality-related information that reflects the app's potential permission needs.
In privacy policies, we extract information specifically related to privacy behaviors.
Previous studies~\cite{yu2017enhancing,alohaly2016better,sathyendra2017identifying} have applied NLP and machine learning techniques to automatically extract specific information.
With the advent of LLMs, which have recently demonstrated remarkable capabilities in NLP tasks, we believe it is viable and more lightweight to use LLMs for specific information extraction from app descriptions and privacy policies.

\begin{figure}[htbp]
    \centering
    \includegraphics[width=0.48\textwidth]{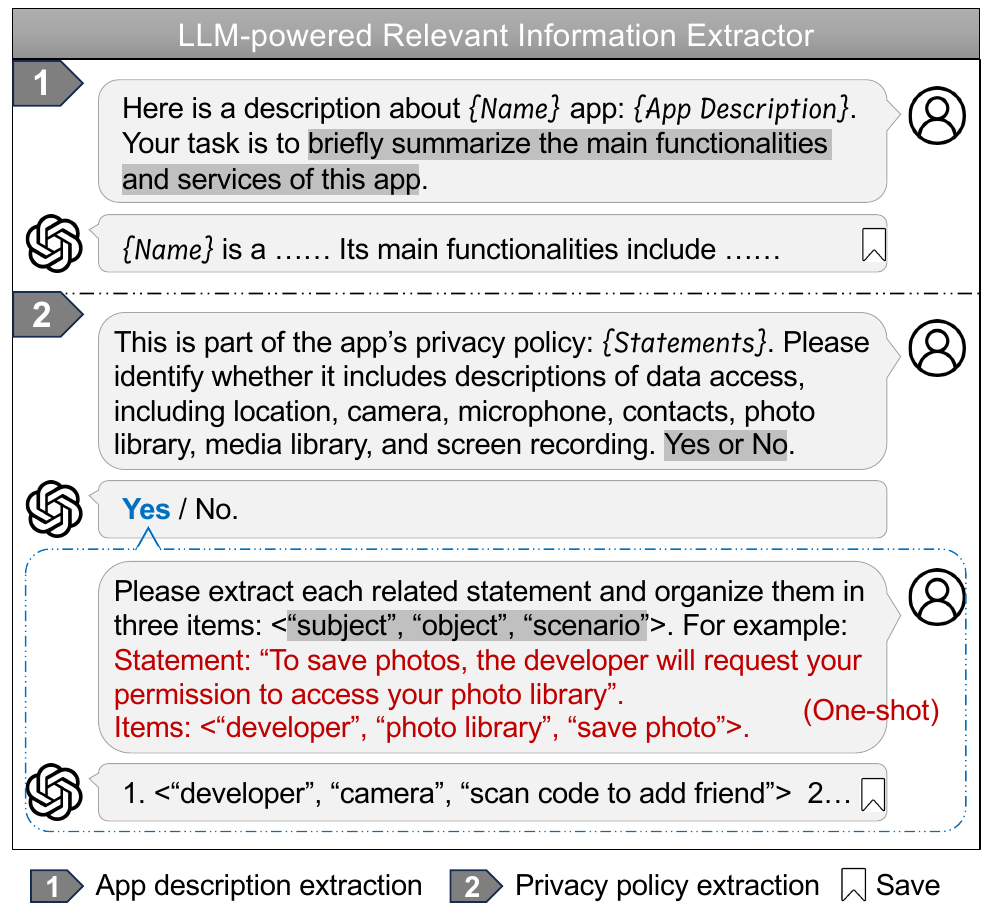}
    \vspace{-0.25in}
    \caption{Examples of prompt engineering of the Extractor.}
    \vspace{-0.1in}
    \label{fig:promptExtractor}
\end{figure}

The simplest way is to directly ask the LLMs to extract the desired information from the original text.
This straightforward approach works well for extracting functionality-related descriptions from app descriptions, as they are typically short and easy to understand. 
However, it is less effective for extracting permission-related statements from privacy policies due to their length, which often exceeds the LLM token limits. 
To address this, we implement a slicing strategy.
First, we divide lengthy privacy policies into smaller, manageable segments that fit within the LLM's token limit. Each segment is then processed sequentially by the LLM to identify statements related to specific data (including location, camera, microphone, contacts, photo library, media library, and screen recording). Only segments containing relevant information are further analyzed to extract details.
This targeted processing ensures we focus on the most pertinent parts and minimize unnecessary processing. 
For simplicity, we represent privacy statements as a series of triples, i.e., \textit{\textless subject, object, scenario\textgreater}, where \textit{subject} indicates the data controller, \textit{object} indicates the data type, and \textit{scenario} describes the specific activities or usage with the data. 
For better understanding, we provide prompt examples in \autoref{fig:promptExtractor}, where a one-shot prompting is used to more effectively extract privacy-related triples.

\subsubsection{LLM-Powered Reasoning}
LLMs have been used to perform program analysis, such as understanding the code constructs and data flows~\cite{li2023hitchhiker,nam2024using}. In this task, LLM was employed to parse the API call stack traces from the sensitive-API tracker.
A well-crafted prompt is crucial when working with LLMs, and
Chain of Thought (COT) prompting has been shown to improve the accuracy of the produced results and the reasoning capabilities of LLMs~\cite{feng2024prompting,nong2024chain}. 
We employ the COT reasoning approach, which guides the LLM through a structured and logical sequence by step-by-step consideration of diverse information sources, including app functionalities, privacy statements, type of data access, and API call stack traces. For instance, the process might start by identifying the specific API caller (class and method names) within the stack trace, which helps determine the usage scenario and data controller. Next, it cross-references the app’s stated functionalities and relevant privacy policy statements to make the inference. 
\autoref{tab:cotptompt} presents particular reasoning examples. In Example \#1, we explain a scenario where the data serves the app's function and is described in the privacy policy. In Example \#2, we explain a scenario where the data serves third-party functionality and is not mentioned in the privacy policy.
The step-by-step reasoning process ensures that the LLM considers all relevant information progressively, leading to more accurate and meaningful interpretations of the data access.

Given the complexity of this task, we strive to keep our COT reasoning prompts concise and efficient. We achieve this by curating specific examples as part of the prompt, tailored to distinct scenarios, i.e., we determine whether the data controller is a common third-party library and provide examples of the purpose of third-party libraries or the app itself. Specifically, to distinguish between first-party and third-party data controllers, we analyze the class names that initiate the API calls, using a reference dataset of third-party library package names created by aggregating known package names from various sources~\cite{zhan2020automated,ma2016libradar,MVN}. When the class name matches a known package in our dataset, we classify it as third-party; otherwise, it's treated as first-party.
This strategy ensures that the information provided is representative rather than exhaustive. Additionally, the number of call stacks processed simultaneously is limited due to a prior deduplication process, while functional descriptions are distilled, and policy statements are streamlined into concise triples. These strategies allow us to maintain a manageable context size while ensuring that the LLM considers all relevant information.

\begin{table}[htbp]
    \centering
    \caption{Examples of COT reasoning about the purpose.}
    \vspace{-0.1in}
    \resizebox{0.48\textwidth}{!}{  
    \begin{tabular}{l|p{8.5cm}} 
    \toprule
     \textbf{Prompt Type} &\textbf{ Instantiation} \\ \hline\hline
     \multirow{2}{*}{Data preparation} &  \texttt{\{Functionalities\}}, 
      \texttt{\{Privacy Statements\}} from the Information Extractor, \newline
    \texttt{\{Call Stack Trace\}}, \texttt{\{Data Type\}} from the dynamic Sensitive-API Tracker.\\\hline
     \multirow{3}{*}{Example Input} & Here is the app: \texttt{\{Functionalities\}}. While running this app, we noticed it accessed \texttt{\{Data Type\}} data through system API. The API call stack trace is as follows: \texttt{\{Call Stack Trace\}}. Its privacy policy states the use of this data as follows: \texttt{\{Privacy Statements\}}. Please infer what the app uses this data for based on these contexts.\\ \hline\hline
     \multirow{4}{*}{\shortstack{Example \#1: \\Chain-of-Thought}} & According to the stack trace, the call is from the \texttt{updateViewWithForecastCardVM} method, suggesting that the location data may be used to update forecast card views. Given that the app supports the functionality of tracking the weather at the current location, and states that it needs location information to provide weather alert pushes, we can infer that the location data is used to get weather and update weather forecast views.\\ \hline
     Example Output &  The location data is used to update the local weather forecast views.\\ \hline\hline
     \multirow{4}{*}{\shortstack{Example \#2: \\Chain-of-Thought}} & According to the stack trace, the call starts from the \texttt{adParam} method, followed by \texttt{requestHotOpenADfinished}, suggesting that the app is requesting advertisements potentially personalized based on the location. This is not listed in the app's main functionalities or privacy statements, and we suspect that this is an undisclosed behavior of the app to tailor ads based on location information.\\ \hline
     Example Output &  The location data is used for targeted advertising.\\
     \bottomrule
    \end{tabular}}
    \label{tab:cotptompt}
\end{table}

\subsection{Domain Clarification}
\label{sec:dom-cla}
In addressing domain clarification, an intuitive method is to present users with a concise and understandable description of each domain's owner, type, and service.
However, a significant challenge arises in acquiring detailed information about the vast array of domains within the extensive mobile network ecosystem.
To address this challenge, our idea is to leverage available resources as much as possible. Specifically, we can use multiple network intelligence sources (e.g., Trackerdb\footnote{An open-source database by Ghostery, offers domain information and maps domains to specific organizations and tracking behaviors. It offers interfaces for developers to integrate the database into non-commercial projects, while vendors can explore commercial partnerships for integration.}~\cite{trackerdb}, Alexa~\cite{alexa}, Netify~\cite{netify}) that offer detailed and reliable information about domains. Additionally, LLMs, with their extensive knowledge base, can be used to identify and generate domain descriptions when direct information is not readily available. As a conservative fallback, we have manual forces, such as analysts, who can manually investigate and classify domains when automated methods fall short.
This multi-tiered strategy can achieve comprehensive coverage in domain clarification.

Based on these insights, we propose a semi-automated pipeline for collecting and presenting domain data. 
As illustrated in \autoref{fig:domain}, this approach involves, first, the systematic collection of domains and their descriptions (including the owner) from reputable network intelligence sources that provide extensive and reliable data about various domains, ensuring the initial database is robust and trustworthy.
The collected data is then meticulously curated to remove duplicates, correct inconsistencies, and refine descriptions where necessary. This curated data is stored in a dedicated allowlist database, which serves as a centralized repository for domain information.
This allowlist database is directly connected to the App Privacy Report. When an app accesses a website, the system immediately queries the database to determine if the domain is listed. If a match is found, the corresponding description is promptly retrieved and included in the App Privacy Report, providing the user with immediate insights into the website being accessed. 
If the domain is not found in the allowlist database, the system seamlessly transitions to the next step: querying an LLM via appropriate prompts. As LLMs are trained on extensive web data, they can attempt to identify and describe the domain. For example, if the domain \texttt{ef-dongfeng.tanx.com} is accessed, the LLM recognizes it as ``Owned by Alibaba Group, part of the Taobao Ad Network Exchange (TANX), providing programmatic advertising services and real-time bidding for online ads'', and this description is stored in the database for future reference and presented to the user in the Report.
In cases where the LLM cannot identify the domain, the pipeline incorporates a manual fallback mechanism where the domain is handed over to an analyst for manual inspection. This ensures no domain is left unclassified, maintaining the comprehensiveness of the database.

This data collection process, although straightforward, is designed for continuous updates and scalability. New domains and updated descriptions can be added regularly, ensuring the database remains current and exhaustive. Automated scripts can periodically scrape data from network intelligence sources, and the LLM can be retrained with new data to improve its accuracy and coverage. Furthermore, when the feature reaches a large user base, crowdsourcing the database maintenance is a viable option.
This pipeline enables users to gain insights into the nature and purpose of the websites accessed by the app, thereby enhancing their understanding and control over their data privacy.

\begin{figure}[htbp]
    \centering
    \includegraphics[width=0.48\textwidth]{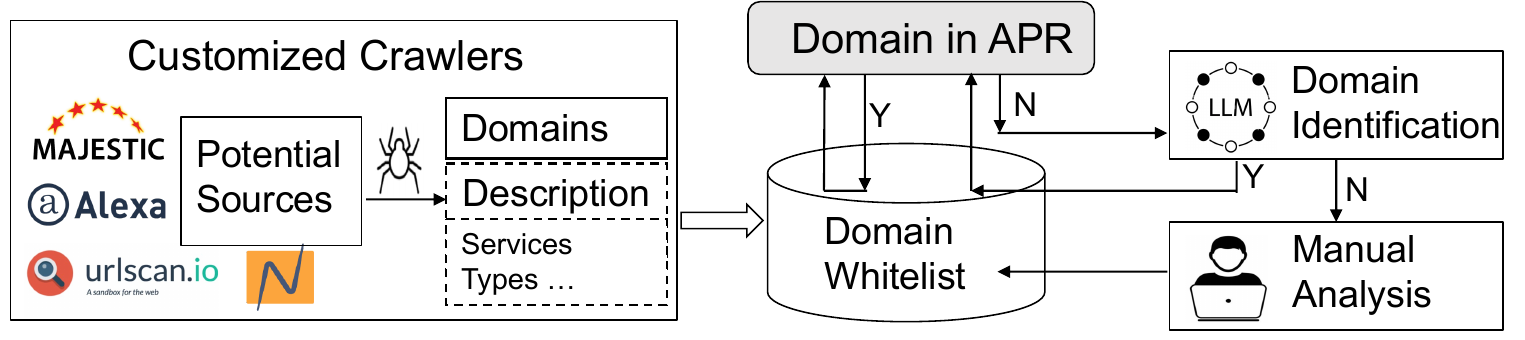}
    \caption{Pipeline of the domain clarification framework (\textit{APR} indicates App Privacy Report).}
    \label{fig:domain}
\end{figure}

\section{Evaluation}
\label{sec:evaluation}
This section reports on the evaluation of the proposed enhancements. Since our enhancements address two key aspects, we evaluate each aspect individually in RQ1 and RQ2.  
Furthermore, we examine whether the enhancements contributed to the overall utility of the App Privacy Report in practice, which is our ultimate objective, in RQ3.

\begin{itemize}[leftmargin=10mm]
    
    \item[RQ1.] How effective is our approach in inferring the purpose of permission usage?
    \item[RQ2.] How effective is our approach in identifying the accessed domain names?
    \item[RQ3.] How useful are our enhancements for providing user privacy insights in the real world?
\end{itemize}

\subsection{RQ1: Effectiveness of Purpose Inference}

\subsubsection{Experimental Setup}
We ran the Purpose Inference framework on an iPhone (iOS version 16.7.5) and a MacBook Pro with an Apple M2 chip and 16-core CPU. We selected an iOS device with a version higher than 15.2 because the App Privacy Report is available from version 15.2 onwards. In experiments, we use manual testing, which triggers privacy-related behaviors more accurately and comprehensively than automated triggering tools such as \textit{NoSmoke}~\cite{nosmoke}.
Manual testing also effectively simulates real-life user-app interactions. We randomly selected 46 apps for the experiment, 2 from each of the 23 categories on the App Store. We also downloaded their app descriptions and privacy policies from the App Store. We used the state-of-the-art \texttt{gpt-4o} model~\cite{gpt-4o} available at the time of the experiment as the backbone for information extraction and purpose reasoning.

\subsubsection{Coverage of Data Access}
We first look at how much data access our sensitive-API tracker could capture. We set up a metric that compares the tracked data with that reported in the App Privacy Report. The underlying assumption is that, since the Privacy Report comes from the official iOS system monitoring, the information in the Report should be accurate and reliable, making it a benchmark for our validation process. 
As a result, for 42 test apps (the other 4 apps failed the test due to crashes or environment-related issues),
we triggered a total of 123 items of sensitive data, an average of 2.93 items of data per app, based on the Privacy Report. 
Our tracker could capture 117 items among them. The missing six included photos (2), contacts (1), screen recording (2), and media library (1). Further manual analysis revealed that the forms of access data for these items were beyond the scope of our current hooks. Such issues can be mitigated by integrating static analysis techniques to expand the scope of dynamic monitoring.

\begin{table*}[htbp]
    \centering
    \caption{Example results of reasoning about the purpose of data access while running the app.}
    \vspace{-0.1in}
    \resizebox{1.0\textwidth}{!}{
    \begin{tabular}{|c|c|c|l|c|}
    \hline
     \textbf{App} &\textbf{Operation} & \textbf{Data Access} & \textbf{Infered Purposes} & \textbf{Result} \\ \hline\hline
      \multirow{5}{*}{Weibo} & \multirow{2}{*}{Post} & Location & Position marking for published posts. & \checkmark\\
        &  & Photo library& To select photos for posts. & \checkmark\\ \cline{2-5}
      &Live  & Camera \& Microphone & For live streaming and video recording. & \checkmark\\ \cline{2-5}
      &\multirow{2}{*}{Click scan icon}    & Camera& To scan QR Code. & \checkmark\\
       &   & Photo library& To select photos, potentially for QR code interaction. & \checkmark\\
          \hline\hline
      \multirow{8}{*}{Jingdong} & Set avatar & Photo library & To select and upload photos, related to profile management and content creation. & \checkmark\\ \cline{2-5}
      & Manage address & Location & To enhance navigation, provide personalized recommendations, and help delivery tracking. & \checkmark\\ \cline{2-5}
      & \multirow{2}{*}{Contact customer service} & Microphone & To enable voice-based interactions, e.g., voice searches and customer service interactions. & \checkmark\\ 
      & & Photo library &  To select photos or videos from the library for sharing or uploading & \checkmark\\ \cline{2-5}
      & \multirow{2}{*}{Click Style Recognition} & Camera & For barcode scanning, AR features, and direct user interactions (e.g., tapping a camera button). & \checkmark\\
      & & Photo library & To check availability and retrieve recent images, possibly for display within the app. & \checkmark\\ \cline{2-5}
      & \multirow{2}{*}{Click scan icon} & Camera & For features like QR code scanning, augmented reality (AR), etc. & \checkmark\\
      & & Photo library & For features like uploading photos and selecting images for products. & \checkmark\\ \cline{2-5}
      &Save product images & Photo library & For features like photo shopping, order verification, and user-generated content.& $\times$ \\
         \hline
    \end{tabular}}
    \label{tab:examplepurpose}
\end{table*}

\subsubsection{Effectiveness of Context Information Extraction}
We also assessed the accuracy and completeness of the LLM-extracted information from privacy policies. To do this, we manually flagged 200 statements related to specific data access from the privacy policies of our test apps and extracted the triples (\textit{\textless subject, object, scenario\textgreater}) from them. Two experienced annotators from the author team independently extracted each statement, with any discrepancies resolved through further discussion. This process yielded 266 triples involving the key data of interest. 
Then, these flagged privacy policies were processed by the LLM to see if it could successfully extract the corresponding triples. The LLM's output was compared against the manually annotated results to measure its accuracy. 
The results showed that the LLM successfully and accurately identified 192 out of 200 statements, containing 257 triples. For the remaining 8 statements (with 9 triples), issues included missing data items or failure to distinguish data controllers. This limitation could potentially be mitigated by fine-tuning the LLM to enhance specificity or by supplementing it with rule-based filters to improve extraction accuracy.
This result highlights the practical utility of using LLMs for targeted information extraction tasks within the context of mobile privacy policies.

\subsubsection{Accuracy of Inferred Purposes}
Next, we evaluate the accuracy of our approach in inferring the purpose of data access from dynamic code behavior and app documentation.
For each test app, we captured call stacks when accessing certain data and analyzed them with context information using LLM-powered reasoning.
The inference results were manually evaluated by two authors with relevant expertise, who considered various factors such as specific user operations, call stacks, and relevant descriptions to make judgments. The key criterion for judgment was that the inferred purpose should be specific, clear (not ambiguous or overly broad), consistent with the user’s current actions, and not conflict with the information suggested by class/method names in the call stack.
Each expert independently reviewed the purpose inference results for each app's data access.

The results showed that each expert individually reported accuracy rates of 91\% and 93\%, respectively, with agreement on 90\% of cases. Thus, we conclude that over 90\% of the inferred purposes were deemed accurate under the experts’ strict criteria. 
It is important to note that the remaining 10\% were not necessarily wrong; often, these inferences were slightly general (e.g., ``providing/enhancing services''), leading to their exclusion under stringent review. 
Upon further analysis, we found that the incorrect or generic inferences primarily resulted from the absence of explicit semantic cues in the call stacks to fully reflect functional details. In these cases, the inferences relied more heavily on the background knowledge derived from app functionalities and privacy statements, which could be somewhat vague or coarse-grained. 

We provide two examples from interactions with Weibo, a popular social media platform often likened to Twitter, and Jingdong, a major e-commerce platform (\autoref{tab:examplepurpose}). 
For each app, the tester performed various operations, some of which triggered access to certain sensitive data, as listed in \autoref{tab:examplepurpose}.
For Weibo, we can see that it is possible to infer the purpose of data usage at a fine-grained level. This was facilitated by the precise naming of classes and methods that clearly reflect the functionality. While, 
for Jingdong, the semantic information from call stacks was less sufficient, leading to more general inferred purposes that covered a broader range of possibilities rather than specifying a particular operation's purpose in detail.  
These results underscore the significant role of source code naming conventions on the purpose inference based on call stack traces: the more descriptive the naming in terms of functionality, the more precise the inferred data access purpose.

Nevertheless, most of the inferred purposes for Jingdong were still deemed correct. This suggests that sometimes even without sufficient call stack information, our method can still derive meaningful inferences by leveraging the app's functionalities and privacy policy statements.
Therefore, we argue that the app description and privacy policy determine the coarse-grained purpose (the lower bound), while the call stack determines the fine-grained purpose (the upper bound), with its precision contingent on the readability and interpretability of the code. 
By incorporating all three data sources, we can cross-verify information and provide broader contexts, thereby maximizing the effectiveness of purpose reasoning.

\subsection{RQ2: Performance of Domain Identification}
\subsubsection{Experimental Setup}
We evaluate performance in two ways. First, since our method relies on LLMs for domain name identification--a process that may extend beyond manual verification--we assess the accuracy of the LLM's identification. Second, we evaluate the domain name coverage provided by our method to ensure comprehensive information for users.
For this evaluation, we recruited two participants (a male and a female) from the previous focus group to collect domain-related data in real-world scenarios.
Before participation, both individuals were fully informed about the experimental procedures, including sharing a subset of domain names with an LLM, and provided explicit consent. Their participation was entirely voluntary and uncompensated.
We instructed them to enable the Privacy Report and use their phones as usual. After two weeks, we examined the domain data from their Privacy Reports and randomly selected 100 domains per participant, totaling 165 unique domains.
Each domain name was then processed through an LLM (\texttt{gpt-4o} model~\cite{gpt-4o}) for identification and description generation, with the outputs manually verified for accuracy. We also look at how many of these domains can be covered by our proposed pipeline.

\subsubsection{Accuracy of LLM's Domain Identification}
We first report on the accuracy of the LLM in identifying domains. Of the 165 domains, the LLM effectively identified and described 160 (for the remaining 5 domains, the LLM indicated a lack of information, and the interaction was closed). Each of these identifications was manually verified, resulting in a 100\% accuracy rate. This suggests the robustness of the LLM's knowledge base in handling domain identification tasks, and our pipeline is feasible.

\subsubsection{Coverage of Domains}
For the 5 domains that the LLM failed to recognize, we conducted a manual exploration and found that they all could be elucidated through network intelligence websites such as \textit{URLScan}. 
Although our test is limited to 165 domains due to manual verification constraints, the fact that all these domains were identified either by LLM or existing knowledge sources highlights the promising effectiveness of our proposed pipeline, which incorporates both the LLM's knowledge and external network intelligence sources to achieve comprehensive domain coverage. With scaling and organizational support, this method could become a highly robust and powerful solution.

\subsection{RQ3: Usefulness of Enhancements}
\subsubsection{Method}
To investigate the utility of our enhancements, we conducted a ``Think Aloud'' (TA) study~\cite{TA}. Think Aloud is a widely used method in user experience testing where participants are encouraged to verbalize their thoughts, feelings, and actions while interacting with the interface~\cite{alhadreti2017intervene,zhao2014impact,cho2019eye}.
This method provides direct insights into the cognitive processes, helping researchers understand user behavior and identify usability issues or cognitive patterns.

\subsubsection{Procedure}
We created visual prototypes of the enhanced interfaces for the Privacy Report, highlighting the addition of the purpose of data access and description of domains (detailed interfaces and processes are available in Appendix~\ref{sec:ta}).
We recruited 11 participants for this evaluation: four returning participants from our initial focus group and seven new iOS users. The returning participants can provide insights on whether the enhancements addressed their previous concerns and expectations, while the new participants can offer fresh, direct feedback on the redesigned interfaces, which can help us optimize the design further.
Before the experiment, all participants signed the consent form and received a demonstration on how to ``think aloud''.
During the TA session, each participant was given a scenario in which they were asked to check specific sections of the App Privacy Report. 
They were encouraged to verbalize their thoughts or feelings as they navigated the pages. 
All participants completed the verbalization process individually.
The verbal response was audio-recorded and researchers took detailed notes. All the data was anonymized and securely stored and explicit informed consent was obtained in advance from the participants following the permission of our institutional Academic Ethics Review Committee.
Each participant was compensated with 20 Chinese yuan ($\approx$ 2.76 dollars).
This remuneration, consistent with our previous focus group study, amounts to 2.3 times the local minimum hourly wage and 1.8 times the median hourly wage, ensuring fair compensation for participants' time and contribution.

\begin{figure}[htbp]
    \centering
    \subfigure{
    \includegraphics[width=0.23\textwidth]{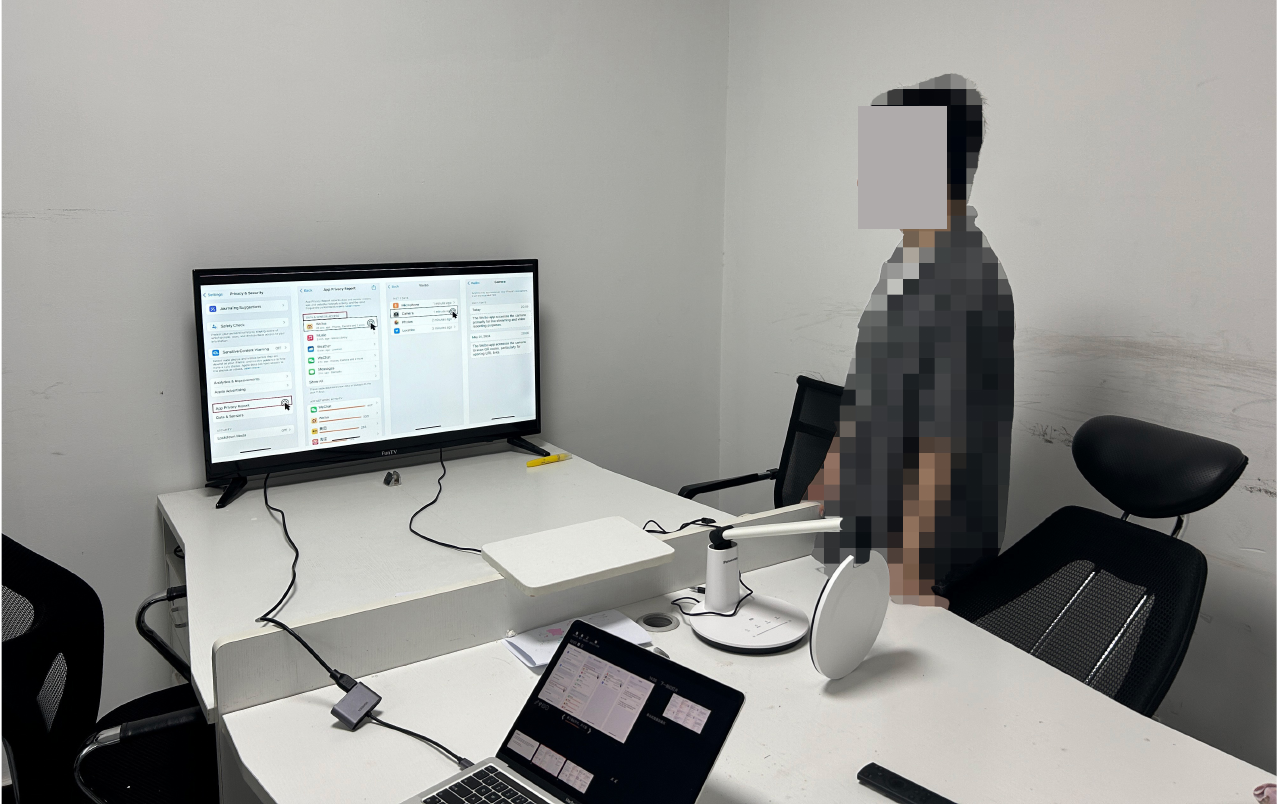}
    \hspace{-0.15in}
    }
    \subfigure{
    \includegraphics[width=0.23\textwidth]{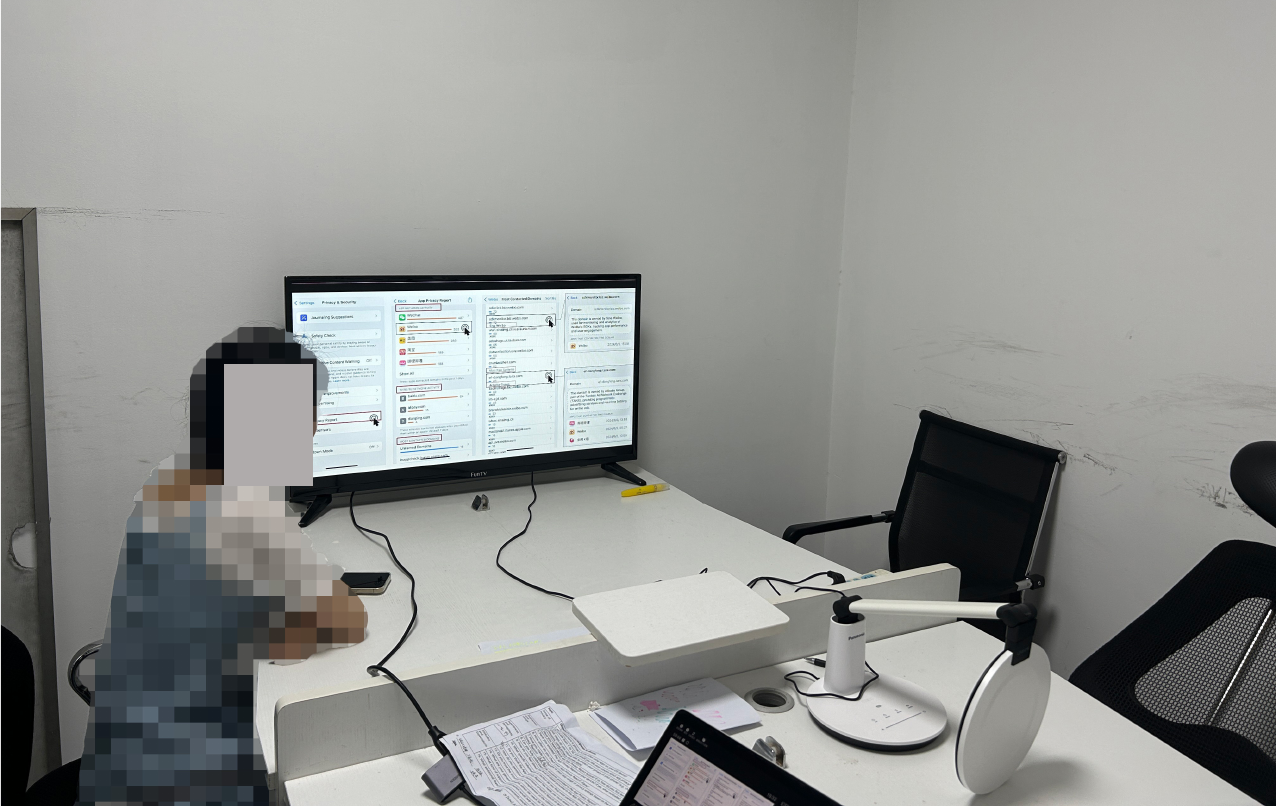}
    }
    \vspace{-0.1in}
    \caption{Two participants (identities obscured) were in their TA sessions.}
    \vspace{-0.05in}
    \label{fig:2}   
\end{figure}

\subsubsection{Results}
Each participant took 9-18 minutes to complete the tasks, with new participants taking slightly longer to read the information on the pages due to their lack of familiarity with the feature.
We have some post-session questions: returning participants were asked about their satisfaction with the new version and whether it addressed concerns with the original, while newcomers were asked about their perceptions of the new version (given their lack of prior knowledge) to evaluate if the improved design met user needs. 
Overall, user feedback was very positive. All participants agreed that the displayed (improved) pages provided sufficient information for them. 
The returning participants confirmed that their concerns were well-addressed by adding the purpose of data access and the domain descriptions. 
As Participant 4 noted, ``Yes, this is exactly what I wanted to see. The new page fulfills what I had envisioned.''
The newcomers expressed satisfaction with both the functionality and content presentation, and they emphasized the significance of the two added elements when asked if they were necessary. 
As Participant 11 commented: ``Of course these elements are important as they provide contextual information about data access purposes and domain connections, which are very useful for me.''
We also asked the newcomers several questions from the previous focus group discussion regarding their perceptions and attitudes toward the displayed feature, to which they responded with predominantly positive feedback. 
When asked about any suggestions, participants offered some suggestions for potential UI enhancements including grouping domains by the owner or by service, folding the descriptions of the purpose and domain, etc. 
Despite these suggestions, participants also stated that the current display was good and acceptable.
As Participant 8 noted, ``I could see that if there were excessive data use, the page might feel crowded; perhaps folding detailed information could help, and I would click to expand when I'm curious about specific accesses. But as it stands, this design is also acceptable.''
Overall, our improved design significantly enhances the usability of the APP Privacy Report feature.

\section{Discussion}

\subsection{Other User Concerns}
\label{sec:concerns}
Beyond the two common issues discussed in this paper, several participants expressed additional concerns. They pointed out that the current App Privacy Report lacks details on what specific data the apps collect or send. 
For instance, while users can see that an app accessed their contacts, they do not know exactly what information was retrieved. Similarly, users might observe that an app contacted multiple external domains but remained unaware of the data sent to these addresses. 
Additionally, a few mentioned that the current Privacy Report does not cover all types of sensitive data, such as calendars and messages are not included. Future updates are expected to include more data types.
Addressing these challenges will require collaborative efforts from app markets, developers, and privacy advocates.

\subsection{Adaptability and Broader Relevance}
Although our study is grounded in the iOS context, the methods we propose are adaptable to other platforms, and the usability issues we observed echo broader trends reported in prior privacy research. For example, Google Play introduced a privacy dashboard~\cite{dashboard} starting with Android 12, which informs users about apps that have accessed their location, camera, and microphone in the past 24 hours (extended to 7 days in Android 13), similar to Apple's App Privacy Report. Yet, this dashboard also lacks clarity on data access purposes and does not include network activity. Since Android apps also use specific system APIs for permissions, our purpose inference method is equally applicable to the Android platform. 
Besides, the issues identified in our focus group study are not unique to the App Privacy Report; they reflect a broad challenge within existing privacy notices, which tend to focus on enumerating data types while offering limited clarity on the specific purposes for which data is used or shared. This shortfall not only undermines user understanding but also fails to meet the transparency requirements under privacy regulations. Therefore, these concerns warrant great attention from privacy professionals across various platforms.

\section{Limitations \& Future Work}

\noindent\textbf{Participant Selection.}
Our recruitment process may have resulted in a sample that is not truly representative of all iOS users. 
We focused on Generation Z participants, who represent a highly active and digitally literate user group in the mobile ecosystem. Their familiarity and frequent interactions with mobile platforms position them well to articulate usability challenges and provide meaningful feedback.
We acknowledge that this demographic focus may limit the generalizability of our findings to other age groups; however, the insights derived from our focus group discussion and Think Aloud evaluation still offer valuable perspectives on the current state of the iOS App Privacy Report.

\vspace{0.05in}
\noindent\textbf{Limitation of LLMs.} 
LLMs are not a panacea and come with inherent limitations. Particularly, they can generate varying results for the same prompt across different runs. The rapidly evolving nature of LLMs might invalidate our designed prompts. 
Regular adaptation and prompt refinement are necessary to ensure compatibility with updated LLM versions, allowing for reliable outputs.

\vspace{0.05in}
\noindent\textbf{Environment Dependency.} 
The purpose inference framework is built on Frida in jailbroken environments, which inherits certain limitations. It cannot operate on apps that apply ``Block Frida Toolkits'' to prevent code instrumentation or those that disable functionality on jailbroken devices. Despite this, our experiment has demonstrated the feasibility of inferring an app's purpose through its API call stack and context information. This could inspire Apple to develop and integrate a similar mechanism in an official capacity.

\vspace{0.05in}
\noindent\textbf{Susceptibility to Code Quality.}
The dynamic analysis relies on the code execution process, so code quality inevitably influences its effectiveness. Coding errors or design flaws can lead to unnecessary API calls, capturing redundant information. For example, we might intercept a location API within a method only to find that the returned location data is unused, resulting in unnecessary capture.
To address this, we can integrate static code analysis to filter out unused API calls before runtime, ensuring a focus on meaningful data usage.
Additionally, our approach relies on semantic features of code, which can be affected by inconsistent or unclear naming conventions, as well as code obfuscation techniques.
Our system could further use context-based heuristics, such as analyzing surrounding code functions, and secondary cues like related API calls or comments to better interpret data access purposes when naming conventions are unclear.

\vspace{0.05in}
\noindent\textbf{Performance and Overhead Considerations.} 
Our enhancements, while dynamically operating at runtime, do not deliver real-time results. Delays arise from logging the call stack and using LLM for purpose inference, as well as domain retrieval and using LLM for domain clarification. The LLM's analysis is the primary time overhead, typically taking a few seconds per interaction. 
Despite this, the time overhead does not affect user experience as the Privacy Report is typically reviewed later. Additionally, instrumentation and domain lookup introduce system overhead. Future applications will require performance optimization and efficient resource management.

\section{Related Work}

\subsection{Assessing the Effectiveness of Privacy Features}
The effectiveness of privacy features in mobile systems has received growing attention, with studies mainly evaluating mechanisms such as privacy policies~\cite{krumayReadabilityPrivacyPolicies2020a,ibdahWhyShouldRead2021a,pan2024trap,pan2024hope}, privacy nutrition labels~\cite{zhang2022usable,koch2022keeping}, App Tracking Transparency (ATT)~\cite{degiulioAskAppNota,mohamed2024attention}.
For example, 
Ibdah et al.~\cite{ibdahWhyShouldRead2021a} examined user attitudes toward privacy policies and reported that even motivated users struggle with technical jargon, leading to disengagement. 
In the context of privacy nutrition labels, Zhang et al.~\cite{zhang2022usable} found that iPhone users frequently misunderstood label content due to unclear terminology and presentation. 
For ATT, Mohamed et al.~\cite{mohamed2024attention} revealed that 59\% of apps utilized dark patterns in ATT prompts to nudge users toward enabling tracking, undermining the intended protections.
These studies consistently highlight challenges in user comprehension, engagement, and trust, revealing limitations in how well current features support informed privacy decisions.
To date, the only study evaluating App Privacy Report feature is Wu et al.~\cite{wutransparency}, which similarly found that users often felt overwhelmed by the volume and technical complexity of network activity data--indicating a risk of information overload despite improved transparency. Their findings align closely with ours. 
However, their work focused on identifying usability limitations, our work, in contrast, goes a step further by proposing feasible enhancements to address them.
In summary, while privacy features like privacy policies, labels, ATT, and App Privacy Report aim to enhance user control and transparency, they often fall short due to issues with readability, accuracy, and user comprehension. These studies collectively emphasize the need for more user-friendly and effective privacy frameworks.

\subsection{Purpose Inference of Permission Use}
To enhance user awareness of sensitive data usage, researchers have developed methods~\cite{pandita2013whyper,qu2014autocog,watanabe2015understanding,wang2017understanding,yang2022describectx,liuMiningAndroidApp,jinWhyAreThey2018} for inferring the purpose behind permission use. WHYPER~\cite{pandita2013whyper} applies NLP techniques to analyze app descriptions, extracting sentences that justify permission requests and establishing a semantic model to determine their relevance. 
CLAP~\cite{liuMiningAndroidApp} employs text mining to recommend explanations for permission requests by aggregating descriptions from similar apps. 
However, these methods rely heavily on textual metadata, limiting their applicability when descriptions are vague, incomplete, or misleading.
Another line of research incorporates static or dynamic analysis for purpose inference.
Wang et al.~\cite{wang2017understanding} extracted static and dynamic features from app code and trained a machine learning classifier to classify purposes. 
DescribeCtx~\cite{yang2022describectx} trained a neural machine translation model using the apps' code and contextual features (via static analysis) to generate descriptive purpose statements. 
While these studies have significantly advanced permission purpose inference, they often suffer from scalability challenges due to high computational costs and the need for curated training datasets. They also face generalization issues, as models trained on specific datasets may struggle to accurately infer permission purposes across diverse app categories and behaviors. 
Our approach builds on these efforts but complements them by introducing a lightweight alternative that combines dynamic instrumentation, call stack analysis, and LLM-based reasoning. 
By avoiding model training and instead leveraging context-aware inference at runtime, our method supports scalable, adaptable, and behavior-driven purpose inference.

\subsection{NLP-Based Privacy Policy Analysis}
Extensive work leverages NLP techniques to improve usability, transparency, and regulatory compliance of privacy policies. These efforts span a range of tasks.
Mazzola et al.~\cite{mazzolaQuestionAnsweringTool2023a} developed a privacy policy QA system that uses NLP models to retrieve relevant policy text in response to natural language questions, enhancing accessibility for non-expert users. Chang et al.~\cite{changAutomatedPersonalizedPrivacy2019a} proposed a personalized extraction framework that tailors policy summaries to individual privacy concerns using keyword classification and rule-based segmentation.
For compliance auditing, PolicyLint~\cite{andowPolicyLintInvestigatingInternal} uses sentence-level NLP and ontology generation to detect internal contradictions in privacy policies. PolicyChecker~\cite{xiangPolicyCheckerAnalyzingGDPR2023} assesses GDPR completeness through semantic role labeling and rule-based checks, while PolicyComp~\cite{zhouPOLICYCOMPCounterpartComparison} extracts and compares privacy policies across similar apps to identify overbroad data collection practices.
Additionally, some studies apply NLP for consistency analysis~\cite{buiConsistencyAnalysisDataUsage2021}, policy classification~\cite{mousavinejadEstablishingStrongBaseline2020a}, or knowledge graph~\cite{cui2023poligraph}.
These approaches primarily employ traditional NLP methods, e.g., rule-based extraction, QA models, or semantic parsing--and focus on auditing, contradiction detection, or summarization. In contrast, our work adopts an LLM-driven approach for extracting relevant privacy statements, using a slicing and prompt-based strategy to address long and complex policy texts. Moreover, our work uses privacy policy content as one of several contextual inputs to support a different task--inferring the purpose behind observed data access behaviors. 
As such, our work extends prior policy-focused analyses by applying LLM-based analysis and repurposing policy content to enhance downstream purpose reasoning tasks.

\section{Conclusion}
While the App Privacy Report marks significant progress in data transparency, its practical impact and potential areas for enhancement remain unclear. 
In this study, we take the first step to examine the practical effectiveness of the iOS App Privacy Report. The focus group discussion highlights user optimism about this feature, but it also reveals existing shortcomings that cannot be ignored. 
Based on user feedback, we propose two promising approaches to enhance the current design. Evaluations based on real user-phone interaction data and the Think Aloud study demonstrate the feasibility and usefulness of our enhancements. Our findings from the focus group discussion and the suggested enhancements can inform the relevant stakeholders to promote better data transparency practices. 
Additionally, our study establishes a research paradigm for systematically evaluating the usability of privacy features, offering a model that integrates user-centered feedback with LLM-driven solutions.

\section*{Acknowledgment}
This work was supported by the National Science Foundation of China (NSFC) under the grant 62172049.

\bibliographystyle{IEEEtran}
\bibliography{cite}
\newpage
\appendices
\label{appendix}

\section{Think Aloud Session}
\label{sec:ta}

\begin{figure*}[htbp]
    \centering
        \subfigure[Provide purpose for each data access (in the grey dotted box on the fourth page).]{
        \includegraphics[width=1\textwidth]{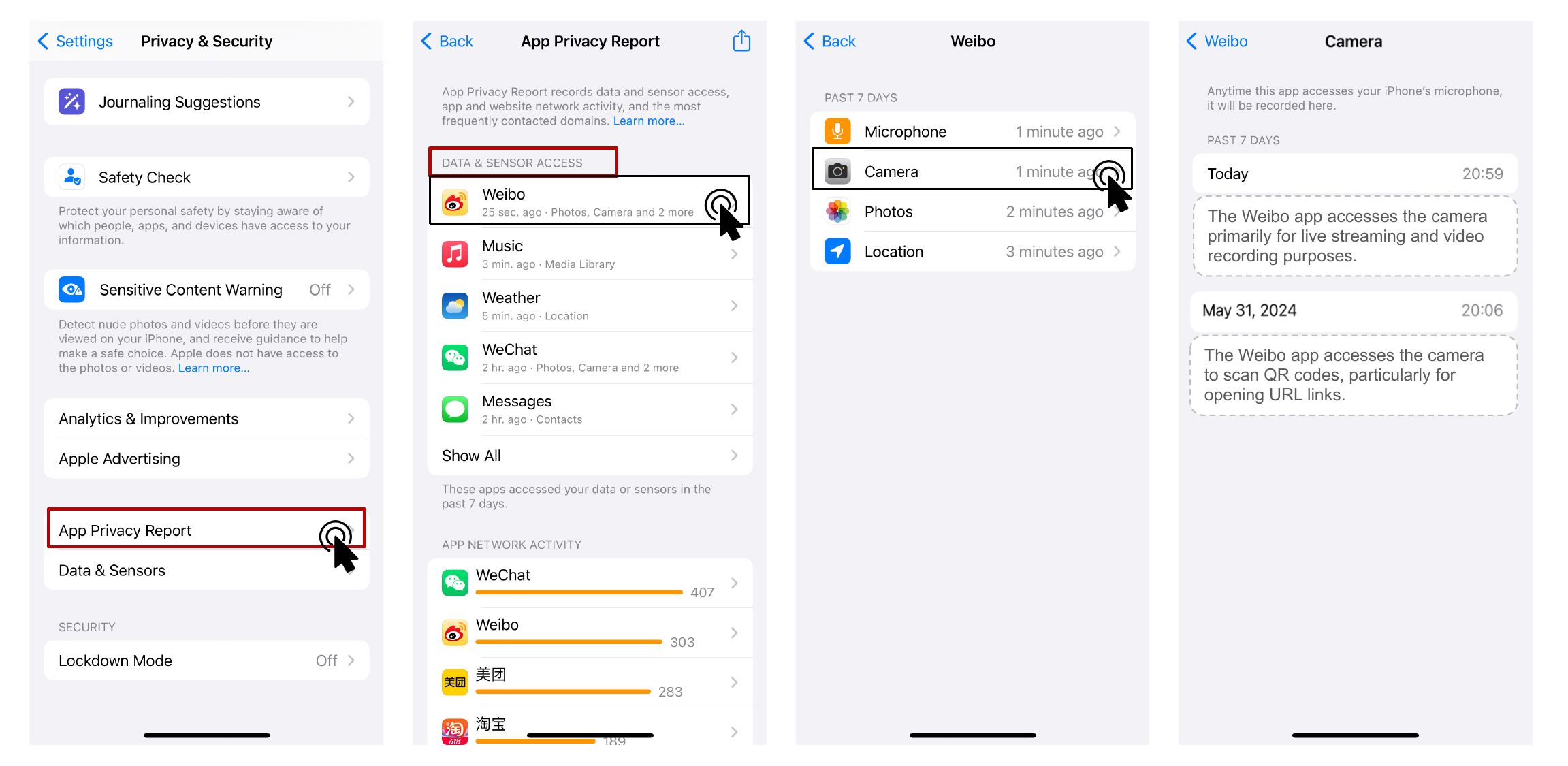}
        \label{fig:ui-a}
        }
        \subfigure[Provide introduction for each domain contact (the organization on the third page and description in the grey dotted box on the fourth page).]{
        \includegraphics[width=1\textwidth]{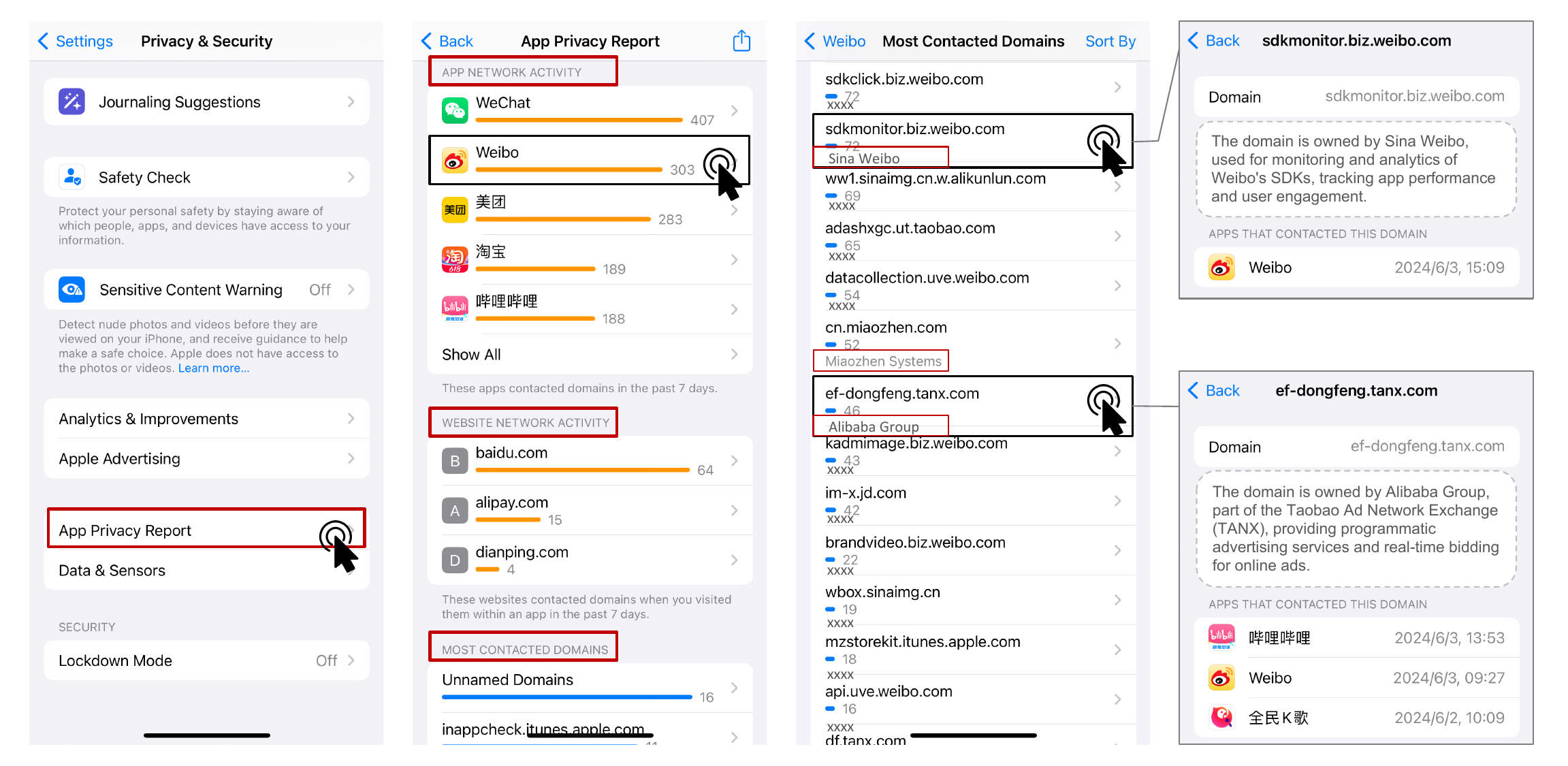}
        \label{fig:ui-b}
        }
        \vspace{-0.1in}
        \caption{The enhanced App Privacy Report interface for data access and domain contact.}
        \label{fig:ui}
\end{figure*}

During the TA session, each participant was given a scenario in which they were asked to review specific sections of the App Privacy Report: 
\textit{You are an iPhone user concerned about privacy. You want to know about the privacy-related behaviors of your recently used apps. So you need to go to Settings \textgreater  Privacy (or Privacy \& Security) \textgreater  App Privacy Report, where you can get a more complete picture of how your apps treat your data.} 

\begin{itemize}
    \item Task 1: Check the data access of your apps.
    \item Task 2: Check the domain connections of your apps.
\end{itemize}

To clarify the experimental details, we provide the visuals of our enhanced App Privacy Report in Figure~\ref{fig:ui}.
Users were instructed to navigate through the interfaces, as if they were viewing it on their own phone. 

In Task 1, users were prompted to check data access information for their apps, as shown in figure~\ref{fig:ui-a}. In the Data \& Sensor Access section, users could view specific data types accessed by each app (e.g., microphone, camera, photos, location) over the past seven days. Selecting a data type, such as the camera for the Weibo app, led users to a detailed view where we introduced the following enhancement, i.e.,
\textbf{Purpose Descriptions:} For each instance of data access, a clear explanation of the purpose is provided within a grey dotted box. For example, Weibo’s camera access is explained as ``primarily for live streaming and video recording purposes'' or ``for scanning QR codes, particularly for opening URL links.''

In Task 2, users were directed to check domain connections under App Network Activity, as guided in figure~\ref{fig:ui-b}. When users select a specific app, like Weibo, they can view a list of domains most frequently contacted by the app. Enhancements in this section are
\textbf{Domain Details:} Each domain entry displays information about the owning organization and the purpose of the connection, presented in a grey-dotted description box on the last page. For instance, ``sdkmonitor.biz.weibo.com'' is identified as owned by Sina Weibo and used for monitoring and analytics of Weibo’s SDKs, tracking app performance and user engagement. 
Additionally, we display the organization’s name on the previous (third) page to give users an immediate understanding of the domain owner, potentially saving them from needing to click through for more details.

During the Think Aloud process, most participants navigated smoothly through the first three pages, understanding the content clearly without any significant questions, suggesting that these pages are well-designed in terms of layout and content presentation. Often, participants spent relatively more time on the fourth page. Returning participants, drawing on prior knowledge, responded positively and felt the content aligned with their previous expectations. 
New participants generally proceeded through the pages with neutral reactions, gradually understanding the functionality as they verbalized their thoughts. Overall, new participants found the enhanced interface more helpful with the added elements, appreciating the improved clarity.

\newpage
\section{Meta-Review}

\subsection{Summary}
This paper considers the recent App-Privacy report provided by Apple's ecosystem. It employs a focus group to learn about people's experiences with the report and what issues they face to determine what needs to be improved about these reports. Based on the focus group results, the authors propose to address two of the issues identified in the app privacy reports. To address these challenges, the authors construct a system that tracks sensitive APIs and extracts information about what the app is doing and why, which they then proceed to evaluate in terms of its effectiveness at its task and follow up with their former participants and new ones to see if the improvements addressed the previous concerns.

\subsection{Scientific Contributions}
\begin{itemize}
\item Provides a Valuable Step Forward in an Established Field
\end{itemize}

\subsection{Reasons for Acceptance}
\begin{enumerate}
\item Provides a Valuable Step Forward in an Established Field. The paper contributes to the research area of user perceptions and usability of privacy dashboards and privacy disclosure mechanisms. It considers a novel setting that may have issues, investigates those issues, identifies some, proposes and executes a (prototype/proof of concept) improvement on those issues, and then re-tests/investigates whether people find those issues less of a concern.
\item The paper proposes two novel enhancements: (I) an automated purpose inference framework that uses large language models (LLMs) to analyze app behavior dynamically, and (II) an allowlist-based domain awareness mechanism that provides users with better insights into app-network interactions. These solutions address real-world concerns and significantly help to enhance user privacy awareness. Additionally, this paper integrates usability considerations into its solutions.
\end{enumerate}

\end{document}